\documentclass[sigconf]{acmart}

\usepackage{xcite}
\usepackage{algorithmic}
\usepackage{graphicx}
\usepackage{textcomp}
\usepackage{xcolor}
\usepackage{enumitem}
\usepackage{multirow}
\usepackage{bigstrut}
\usepackage[skip=-3pt]{subcaption}
\usepackage[skip=0pt]{caption}
\usepackage[ruled]{algorithm2e}

\renewcommand\footnotetextcopyrightpermission[1]{} 
\makeatletter
\def\subsubsection{\@startsection{subsubsection}{3}%
  \z@{.3\linespacing\@plus.7\linespacing}{.1\linespacing}%
  {\normalfont\itshape}}
\makeatother

\begin{document}

\title{Probabilistic Metric Learning with Adaptive Margin for Top-K Recommendation
}

\author{Chen Ma}
\authornote{Work done as interns at Huawei Noah's Ark Lab Montreal Research Center.}
\affiliation{%
  \institution{McGill University}
}
\email{chen.ma2@mail.mcgill.ca}

\author{Liheng Ma}
\authornotemark[1]
\affiliation{%
  \institution{McGill University}
}
\email{liheng.ma@mail.mcgill.ca}

\author{Yingxue Zhang}
\affiliation{%
  \institution{Huawei Noah's Ark Lab Montreal}
}
\email{yingxue.zhang@huawei.com}

\author{Ruiming Tang}
\affiliation{%
  \institution{Huawei Noah's Ark Lab}
}
\email{tangruiming@huawei.com}

\author{Xue Liu}
\affiliation{%
  \institution{McGill University}
}
\email{xueliu@cs.mcgill.ca}

\author{Mark Coates}
\affiliation{%
  \institution{McGill University}
}
\email{mark.coates@mcgill.ca}

\begin{abstract}
Personalized recommender systems are playing an increasingly important role as more content and services become available and users struggle to identify what might interest them. Although matrix factorization and deep learning based methods have proved effective in user preference modeling, they violate the triangle inequality and fail to capture fine-grained preference information. To tackle this, we develop a distance-based recommendation model with several novel aspects: (i) each user and item are parameterized by Gaussian distributions to capture the learning uncertainties; (ii) an adaptive margin generation scheme is proposed to generate the margins regarding different training triplets; (iii) explicit user-user/item-item similarity modeling is incorporated in the objective function. The Wasserstein distance is employed to determine preferences because it obeys the triangle inequality and can measure the distance between probabilistic distributions. Via a comparison using five real-world datasets with state-of-the-art methods, the proposed model outperforms the best existing models by 4-22\% in terms of recall@K on Top-K recommendation.
\end{abstract}

\maketitle

\section{Introduction}
Internet users can easily access an increasingly vast number of online
products and services, and it is becoming very difficult for users to identify
the items that will appeal to them out of a plethora of candidates.
To reduce information overload and to satisfy the diverse needs of
users, personalized recommender systems have emerged and they are
beginning to play an important role in modern society. These systems
can provide personalized experiences, serve huge service demands, and
benefit both the user-side and supply-side. They
can: (i) help users easily discover products that are likely to
interest them; and (ii) create opportunities for product and service providers to better serve customers and to increase
revenue.

In all kinds of recommender systems, modeling the user-item
interaction lies at the core. There are two common ways used in recent
recommendation models to infer the user preference: matrix factorization (MF) and
multi-layer perceptrons (MLPs). MF-based methods
(e.g.,~\cite{DBLP:conf/icdm/HuKV08,DBLP:conf/uai/RendleFGS09}) apply
the inner product between latent factors of users and items to predict
the user preferences for different items. The latent factors strive to
depict the user-item relationships in the latent space. In contrast,
MLP-based methods
(e.g.,~\cite{DBLP:conf/www/HeLZNHC17,DBLP:conf/recsys/CovingtonAS16})
adopt (deep) neural networks to learn non-linear user-item
relationships, which can generate better latent feature combinations
between the embeddings of users and
items~\cite{DBLP:conf/www/HeLZNHC17}.

However, both MF-based and MLP-based methods violate the triangle
inequality~\cite{DBLP:conf/nips/Shrivastava014}, and as a result may fail to
capture the fine-grained preference
information~\cite{DBLP:conf/www/HsiehYCLBE17}. As a concrete example
in~\cite{DBLP:conf/icdm/ParkKXY18}, if a user accessed two items, MF
or MLP-based methods will put both items close to the user, but will
not necessarily put these two items close to each other, even if they
share similar properties.

To address the limitations of MF and MLP-based methods, metric (distance) learning approaches have been utilized in the recommendation model~\cite{DBLP:conf/www/HsiehYCLBE17,DBLP:conf/icdm/ParkKXY18,DBLP:conf/www/TayTH18,DBLP:journals/corr/LiZZQZHH20}, as the distance naturally satisfies the triangle inequality. These techniques project users and items into a low-dimensional metric space, where the user preference is measured by the distance to items. Specifically, CML~\cite{DBLP:conf/www/HsiehYCLBE17} and LRML~\cite{DBLP:conf/www/TayTH18} are two representative models. CML minimizes the Euclidean distance between users and their accessed items, which facilitates user-user/item-item similarity learning. LRML incorporates a memory network to introduce additional capacity to learn relations between users and items in the metric space.

\begin{figure}[ht]
    \centering
    \includegraphics[width=0.45\textwidth]{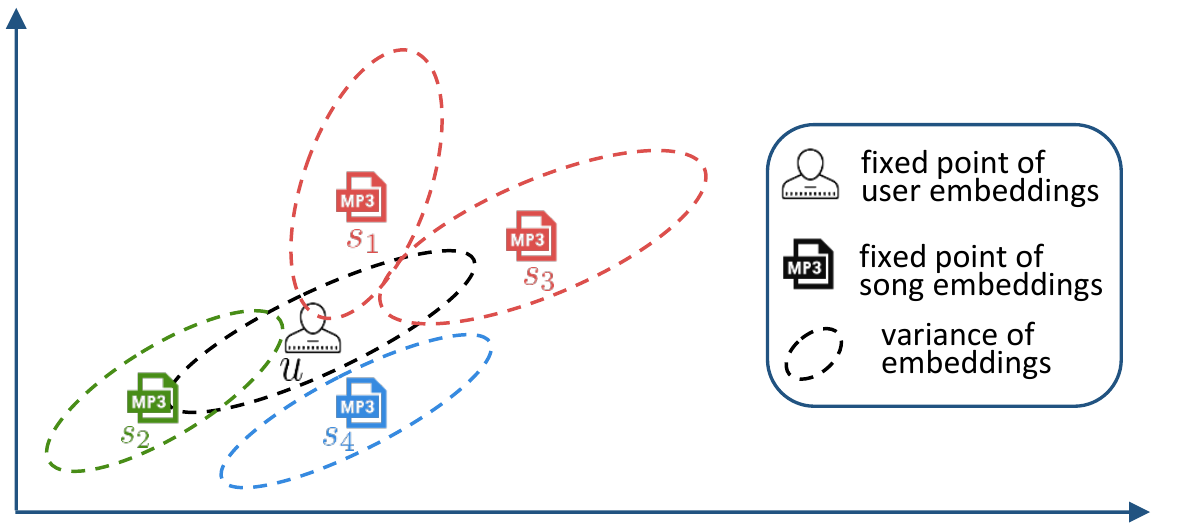}
    \caption{A motivating example of handling uncertainties of learned embeddings.}
    \label{fig:example}
\vspace{-0.5cm}
\end{figure}

Although existing distance-based methods have achieved satisfactory
results, we argue that there are still several avenues for enhancing
performance. First, previous distance-based
methods~\cite{DBLP:conf/www/HsiehYCLBE17,DBLP:conf/icdm/ParkKXY18,DBLP:conf/www/TayTH18,DBLP:journals/corr/LiZZQZHH20}
learn the user and item embeddings in a deterministic manner without
handling the uncertainty. Relying solely on the learned deterministic
embeddings may lead to an inaccurate understanding of user
preferences. A motivating example is shown in
Figure~\ref{fig:example}. After having accessed two songs $ s_1 $ and
$ s_2 $ with different genres, the user $ u $ may be placed between of
$ s_1 $ and $ s_2 $. If we only consider deterministic embeddings,
$ s_4 $ should be a good candidate. But if we consider the embeddings
from a probabilistic perspective, $ s_3 $ can be a better
recommendation and it has the same genre as $ s_1 $. Second, most of
the existing methods~
\cite{DBLP:conf/www/HsiehYCLBE17,DBLP:conf/www/TayTH18,DBLP:conf/icdm/ParkKXY18}
adopt the margin ranking loss (hinge loss) with a fixed margin as the
hyper-parameter. We argue that the margin value should be adaptive and
relevant to corresponding training samples. Furthermore, different
training phases may need different magnitudes of margin values. Setting
a fixed value may not be an optimal solution. Third, previous
distance-based methods~
\cite{DBLP:conf/www/HsiehYCLBE17,DBLP:conf/www/TayTH18,DBLP:conf/icdm/ParkKXY18,DBLP:journals/corr/LiZZQZHH20}
do not explicitly model user-user and item-item relationships.
Closely-related users are very likely to share the same interests, and
if two items have similar attributes it is likely that a user will
favour both. When inferring a user's preferences, we should explicitly
take into account the user-user and item-item similarities.

To address the shortcomings, we propose a Probabilistic Metric
Learning model with an Adaptive Margin (PMLAM) for Top-K
recommendation. PMLAM consists of three major components: 1) a
user-item interaction module, 2) an adaptive margin generation module,
and 3) a user-user/item-item relation modeling module. To capture the
uncertainties in the learned user and item embeddings, each user or
item is parameterized with one Gaussian distribution, where the
distribution related parameters are learned by our model. In the
user-item interaction module, we adopt the Wasserstein distance to
measure the distances between users and items, thus not only taking
into account the means but also the uncertainties. In the adaptive
margin generation module, we model the learning of adaptive margins as
a bilevel (inner and outer) optimization
problem~\cite{DBLP:journals/tec/SinhaMD18}, where we build a proxy
function to explicitly link the learning of margin related parameters
with the outer objective function. In the user-user and item-item
relation modeling module, we incorporate two margin ranking losses
with adaptive margins for user-pairs and item-pairs, respectively, to
explicitly encourage similar users or items to be mapped closer to one
another in the latent space. We extensively evaluate our model by
comparing with many state-of-the-art methods, using two performance
metrics on five real-world datasets. The experimental results not only
demonstrate the improvements of our model over other baselines but
also show the effectiveness of the proposed modules.

To summarize, the major contributions of this paper are:
\begin{itemize}[leftmargin=*]
\item To capture the uncertainties in the learned user/item
  embeddings, we represent each user and item as a Gaussian distribution. The Wasserstein distance is leveraged to measure the user preference for items while simultaneously considering the uncertainty.
\item To generate an adaptive margin, we cast margin generation as a bilevel optimization problem, where a proxy function is built to explicitly update the margin generation related parameters.
\item To explicitly model the user-user and item-item relationships, we apply two margin ranking losses with adaptive margins to force similar users and items to map closer to one another in the latent space.
\item Experiments on five real-world datasets show that the proposed PMLAM model significantly outperforms the state-of-the-art methods for the Top-K recommendation task.
\end{itemize}

\section{Related Work}
In this section we summarize and discuss work that is related to our proposed top-K recommendation model.

In many real-world recommendation scenarios, user implicit
data~\cite{DBLP:conf/kdd/WangWY15,DBLP:conf/kdd/MaKL19}, e.g., clicking
history, is more common than explicit
feedback~\cite{DBLP:conf/icml/SalakhutdinovMH07} such as user ratings.
The implicit feedback setting, also called one-class collaborative
filtering (OCCF)~\cite{DBLP:conf/icdm/PanZCLLSY08}, arises when only
positive samples are available. To tackle this challenging problem,
effective methods have been proposed.

\textbf{Matrix Factorization-based Methods}. Popularized by the
Netflix prize competition, matrix factorization (MF) based methods
have become a prominent solution for personalized
recommendation~\cite{DBLP:journals/computer/KorenBV09}. In~\cite{DBLP:conf/icdm/HuKV08}, Hu et
al.\ propose a weighted regularized matrix factorization (WRMF) model
to treat all the missing data as negative samples, while heuristically
assigning
confidence weights to positive samples.
Rendle et al.\ adopt a different approach in~\cite{DBLP:conf/uai/RendleFGS09},
proposing a pair-wise ranking objective (Bayesian personalized ranking)
to model the pair-wise relationships between positive items and
negative items for each user, where the negative samples are randomly sampled from
the unobserved feedback. To allow unobserved
items to have varying degrees of importance, He et al.\ in~\cite{DBLP:conf/sigir/HeZKC16} propose to weight
the missing data based on item popularity, demonstrating improved
performance compared to WRMF.

\textbf{Multi-layer Perceptron-based Methods}.
Due to their ability to learn more complex non-linear relationships
between users and items, (deep)
neural networks have been a great success in the domain of recommender
systems. He et al.\ in~\cite{DBLP:conf/www/HeLZNHC17} propose a neural
network-based collaborative filtering model, where a multi-layer
perceptron is used to learn the non-linear user-item interactions.
In~\cite{DBLP:conf/wsdm/WuDZE16,DBLP:conf/cikm/MaZWL18,DBLP:conf/wsdm/MaKWWL19},
(denoising) autoencoders are employed to learn the user or item hidden
representations from user implicit feedback. Autoencoder approaches
can be shown to be generalizations of many of the MF
methods~\cite{DBLP:conf/wsdm/WuDZE16}. In
\cite{DBLP:conf/ijcai/XueDZHC17,DBLP:conf/ijcai/GuoTYLH17},
conventional matrix factorization and factorization machine methods
benefit from the representation ability of deep neural networks for
learning either the user-item relationships or the interactions with
side information. Graph neural networks (GNNs) have recently been incorporated in recommendation algorithms because they can learn and model relationships between entities~\cite{DBLP:conf/sigir/Wang0WFC19,DBLP:conf/icdm/SunZMCGTH19,DBLP:conf/aaai/MaMZSLC20}.

\textbf{Distance-based Methods}. Due to their capacity to measure the
distance between users and items, distance-based methods have been
successfully applied in Top-K recommendation.
In~\cite{DBLP:conf/www/HsiehYCLBE17}, Hsieh et al.\ propose to compute
the Euclidean distance between users and items for capturing
fine-grained user preference. In~\cite{DBLP:conf/www/TayTH18}, Tay et
al.\ adopt a memory network~\cite{DBLP:conf/nips/SukhbaatarSWF15} to
explicitly store the user preference in external memories. Park et al.\ in \cite{DBLP:conf/icdm/ParkKXY18} apply a translation emb  edding to capture more complex relations between users and items, where the translation embedding is learned from the neighborhood information of users and items.
In~\cite{DBLP:conf/recsys/HeKM17}, He et al.\ apply a distance metric to
capture how the user interest shifts across sequential user-item interactions. In~\cite{DBLP:journals/corr/LiZZQZHH20}, Li et al.\ propose to measure the trilateral relationship from both the user-centric and item-centric perspectives and learn adaptive margins for the central user and positive item.


Our proposed recommendation model is different in key ways from  all of the methods identified above. 
In contrast to the matrix factorization~\cite{DBLP:conf/icdm/HuKV08, DBLP:conf/uai/RendleFGS09,DBLP:conf/sigir/HeZKC16} and neural network methods~\cite{DBLP:conf/www/HeLZNHC17,DBLP:conf/wsdm/WuDZE16,DBLP:conf/cikm/MaZWL18,DBLP:conf/wsdm/MaKWWL19,DBLP:conf/ijcai/XueDZHC17,DBLP:conf/ijcai/GuoTYLH17,DBLP:conf/sigir/Wang0WFC19}, we employ the Wasserstein distance that obeys the triangle inequality. This is important for ensuring that users with similar interaction histories are mapped close together in the latent space.
In contrast to most of the prior distance-based
approaches,~\cite{DBLP:conf/www/HsiehYCLBE17,DBLP:conf/www/TayTH18,DBLP:conf/icdm/ParkKXY18,DBLP:journals/corr/LiZZQZHH20}, we employ parameterized Gaussian distributions to represent each user and item in order to capture the uncertainties of learned user preferences and item properties. Moreover, we formulate a bilevel optimization problem and incorporate a neural network to generate adaptive margins for the commonly applied margin ranking loss function.



\section{Problem Formulation}
The recommendation task considered in this paper takes as input the user implicit feedback. For each user $ i $, the user preference data is represented by a set that includes the items she preferred, e.g., $ \mathcal{D}_{i}=\{ I_1,...,I_j,...,I_{|\mathcal{D}_{i}|} \} $, where $ I_j $ is an item index in the dataset. 
The top-$ K $ recommendation task in this paper is formulated as: given the training item set $ \mathcal{S}_{i} $, and the non-empty test item set $ \mathcal{T}_{i} $ (requiring that $ \mathcal{S}_{i} \cup \mathcal{T}_{i} = \mathcal{D}_{i} $ and $ \mathcal{S}_{i} \cap \mathcal{T}_{i} = \emptyset $) of user $ i $, the model must recommend an ordered set of items $ \mathcal{X}_{i} $ such that $ |\mathcal{X}_{i}| \leq K $ and $ \mathcal{X}_{i} \cap \mathcal{S}_{i} = \emptyset $. Then the recommendation quality is evaluated by a matching score between $ \mathcal{T}_{i} $ and $ \mathcal{X}_{i} $, such as Recall@$K$.

\section{Methodology}
\begin{figure*}[htb]
    \centering
    \includegraphics[width=\linewidth]{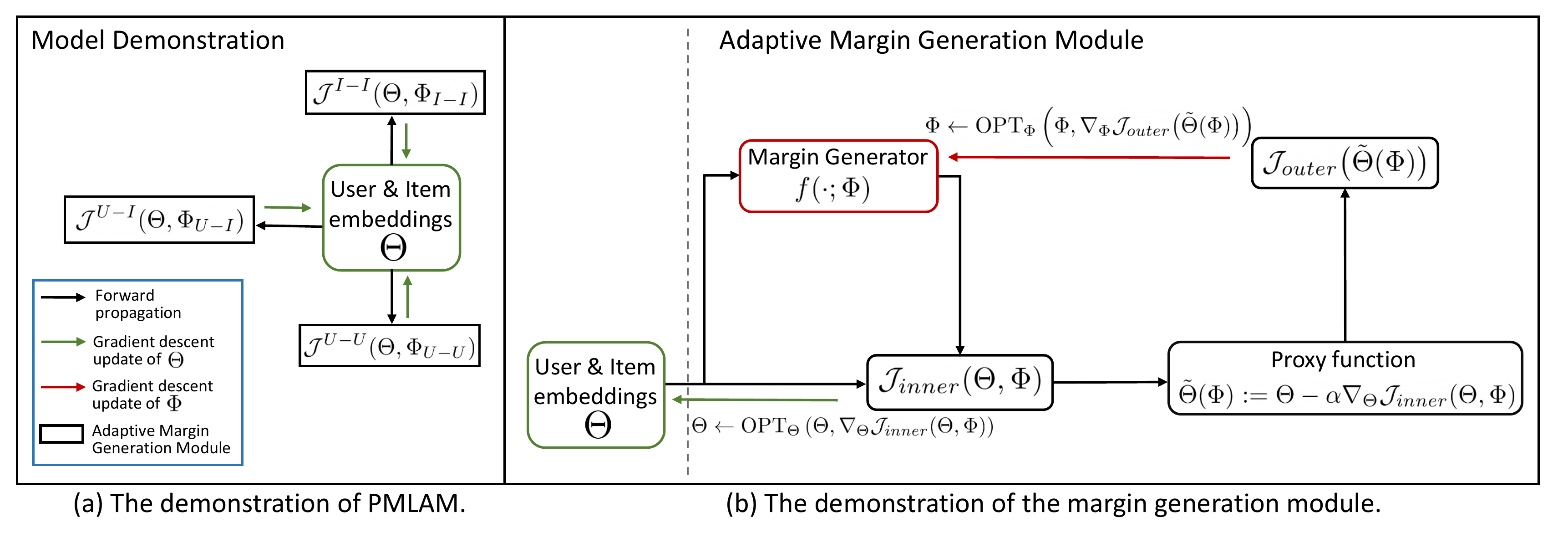}
    \caption{The demonstration of the proposed model. $\mathcal{J}^{U-I}$ denotes the combined optimization regarding $\mathcal{J}_{inner}^{U-I}$ and $\mathcal{J}_{outer}^{U-I}$. $\mathcal{J}^{U-U}$ and $\mathcal{J}^{I-I}$ follow the same manner with $\mathcal{J}^{U-I}$.}
    \label{fig:whole_model}
\vspace{-0.4cm}
\end{figure*}

In this section, we present the proposed model shown in
Fig.~\ref{fig:whole_model}. We first introduce the user-item interaction
module, which captures the user-item interactions by calculating the
Wasserstein distance between users' and items' distributions. Then we
describe the adaptive margin generation module, which generates
adaptive margins during the training process. Next, we present the
user-user and item-item relation modeling module. Lastly, we specify
the objective function and explain the training process of the proposed model.

\subsection{Wasserstein Distance for Interactions} \label{sec:wass}
Previous works~\cite{DBLP:conf/www/HsiehYCLBE17,DBLP:conf/www/TayTH18}
use the user and item embeddings in a deterministic manner and do not
measure or learn the uncertainties of user
preferences and item properties. Motivated by probabilistic matrix factorization (PMF)~\cite{DBLP:conf/nips/SalakhutdinovM07}, we represent each user or item as a single Gaussian distribution. In
contrast to PMF, which applies Gaussian priors on user and item embeddings,
users and items in our model are parameterized by Gaussian
distributions, where the means $ \mu $ and covariances $ \Sigma $ are directly learned.
Specifically, the latent factors of user $ i $ and item $ j $ are represented as: 
\begin{equation}
\begin{aligned}
    \mathbf{u}_i & \sim \mathcal{N}(\mathbf{\mu}^{(U)}_{i}, \mathbf{\Sigma}^{(U)}_{i}) \,, \\
    \mathbf{v}_j & \sim \mathcal{N}(\mathbf{\mu}^{(I)}_{j}, \mathbf{\Sigma}^{(I)}_{j}) \,.
\end{aligned}
\label{eq:gauss_distribution}
\end{equation}
Here $ \mathbf{\mu}^{(U)}_{i} $ and $ \mathbf{\Sigma}^{(U)}_{i} $ are
the learned mean vector and covariance matrix of user $ i $, respectively; $ \mathbf{\mu}^{(I)}_{j} $ and $\mathbf{\Sigma}^{(I)}_{j} $ are the learned mean vector and covariance matrix of item $j$. To limit the complexity of the model
and reduce the computational overhead, we assume that the embedding
dimensions are uncorrelated. Thus, $\mathbf{\Sigma}$ is a diagonal
covariance matrix that can be represented as a vector. Specifically,
$ \mathbf{\mu} \in \mathbb{R}^h $ and
$ \mathbf{\Sigma} \in \mathbb{R}^h $, where $ h $ is the dimension of
the latent space.

Widely used distance metrics for deterministic embeddings, like the Euclidean distance, do not properly measure the distance between distributions. Since users and items are represented by probabilistic distributions, we need a distance measure between
distributions. Among the commonly used distance metric
between distributions, we adopt the Wasserstein distance to measure
the user preference for an item. The reasons are twofold: i) the Wasserstein
distance satisfies all the properties a distance should have; and ii) the
Wasserstein distance has a simple form when calculating the distance
between Gaussian distributions~\cite{DBLP:conf/nips/MallastoF17}.
Formally, the $ p $-th Wasserstein distance between two probability
measures $\mu$ and $\nu$ on a Polish metric
space~\cite{srivastava2008course} $(\mathcal{X}, d)$ is defined~\cite{Givens_1984}:
\[
\mathcal{W}_{p}(\mu, \nu) := \Big(\inf_{J \in \mathcal{J}(\mu,\nu)} \int_{\mathcal{X}\times \mathcal{X}} d(x, y)^p dJ(x,y) \Big)^{\frac{1}{p}} \,,
\label{eq:W_distance}
\]
where $d(\cdot, \cdot)^p$ is an arbitrary distance with $p^{th}$
moment~\cite{casella2002statistical} for a deterministic variable,
$p \in [1, +\infty)$; and $\mathcal{J}(\mu, \nu)$ denotes the set of
all measures $J$ on $\mathcal{X}\times \mathcal{X}$ which admit $\mu$
and $\nu$ as marginals. When $ p \geq 1 $, the $ p $-th Wasserstein
distance preserves all properties of a metric, including both symmetry
and the triangle inequality.

The calculation of the general Wasserstein distance is computation-intensive~\cite{DBLP:conf/uai/XieWWZ19}. To reduce the
computational cost, we use Gaussian distributions for the latent
representations of users and items. Then when $p = 2$, the $ 2 $-nd
Wasserstein distance (abbreviated as $ \mathcal{W}_2 $) has a closed
form solution, thus making the calculation process much faster. Specifically, we
have the following formula to calculate the $ \mathcal{W}_2 $ distance
between user $ i $ and item
$ j $~\cite{Givens_1984}: \begin{equation} \begin{aligned}
    &\mathcal{W}_2 \left(\mathcal{N}(\mathbf{\mu}^{(U)}_{i}, \mathbf{\Sigma}^{(U)}_{i}), \, \mathcal{N}(\mathbf{\mu}^{(I)}_{j}, \mathbf{\Sigma}^{(I)}_{j}) \right)^{2} \\
                &= ||\mathbf{\mu}^{(U)}_{i} - \mathbf{\mu}^{(I)}_{j}||_{2}^{2} + \mathrm{trace} \bigg(\mathbf{\Sigma}^{(U)}_{i} + \mathbf{\Sigma}^{(I)}_{j} -2 \Big( (\mathbf{\Sigma}^{(U)}_{i})^{\frac{1}{2}} \mathbf{\Sigma}^{(I)}_{j} (\mathbf{\Sigma}^{(U)}_{i})^{\frac{1}{2}} \Big)^{\frac{1}{2}} \bigg)\,.
\label{eq:gauss_dist}
\end{aligned} 
\end{equation}

In our setting, we focus on diagonal covariance matrices, thus
$ \mathbf{\Sigma}^{(U)}_{i} \mathbf{\Sigma}^{(I)}_{j} =
\mathbf{\Sigma}^{(I)}_{j} \mathbf{\Sigma}^{(U)}_{i}$. For simplicity,
we use $ \mathcal{W}_2(i, j)^{2} $ to denote the left hand side of
Eq.~\ref{eq:gauss_dist}. Then Eq.~\ref{eq:gauss_dist} can be
simplified as:
\begin{equation}
    \mathcal{W}_2(i, j)^{2} = ||\mathbf{\mu}^{(U)}_{i} - \mathbf{\mu}^{(I)}_{j}||_{2}^{2} + || (\mathbf{\Sigma}^{(U)}_{i})^{\frac{1}{2}} - (\mathbf{\Sigma}^{(I)}_{j})^{\frac{1}{2}} ||^{2}_{2} \,.
\end{equation}
According to the above equation, the time complexity of calculating $ \mathcal{W}_2 $ distance between the latent representations of users and items is linear with the embedding dimension.

\subsection{Adaptive Margin in Margin Ranking Loss} \label{sec:adaptive_margin}
To learn the distance-based model, most of the existing
works~\cite{DBLP:conf/www/HsiehYCLBE17,DBLP:conf/www/TayTH18} apply
the margin ranking loss to measure the user preference difference
between positive items and negative items. Specifically, the margin
ranking loss makes sure the distance between a user and a positive
item is less than the distance between the user and a negative item by
a fixed margin $ m > 0$. The loss function is:
\begin{equation}
    \mathcal{L}_{Fix}(i, j, k;\Theta) = 
    [d(i, j; \Theta)^{2} - d(i, k; \Theta)^{2} + m]_{+} \,,
\label{eq:fixed_margin}
\end{equation}
where $ j \in \mathcal{S}_i $ is an item that user $ i $ has accessed, and
$ k \not \in \mathcal{S}_i$ is a randomly sampled item treated as
the negative example, and $ [z]_{+} = \max(z, 0) $. Thus, $ (i, j, k) $ represents a training
triplet.


The safe margin $m$ in the margin ranking loss is a crucial
hyper-parameter that has a major impact on the model performance. A
fixed margin value may not achieve satisfactory performance. First,
using a fixed value does not allow for adaptation to distinguish
the properties of the training triplets. For example, some users have
broad interests, so the margins for these users should not be so large
as to make potential preferred items too far from the user. Other
users have very focused interests, and it is desirable to have a
larger margin to avoid recommending items that are not directly within
the focus. Second, in different training phases, the model may need
different magnitudes of margins. For instance, in the early stage of
training, the model is not reliable enough to make strong predictions
on user preferences, and thus imposing a large margin risk pushing
potentially positive items too far from a user. Third, to achieve
satisfactory performance, the selection of a fixed margin involves tedious
hyper-parameter tuning. Based on these considerations, we conclude
that setting a fixed margin value for all training triplets may limit the model expressiveness.

To address the problems outlined above, we propose an adaptive margin
generation scheme which generates margins according to the training
triplets. Formally,
we formulate the margin ranking loss with an adaptive margin as:
\begin{equation}
    \mathcal{L}_{Ada}(i, j, k; \Theta, \Phi) = 
    [d(i, j; \Theta)^{2} - d(i, k; \Theta)^{2} + f(i, j, k; \Phi)]_{+} \,.
\label{eq:adaptive_margin}
\end{equation}
Here $ f(i, j, k; \Phi) $ is a function that generates the specific
margin based on the corresponding user and item embeddings and $ \Phi $ is
the learnable set of parameters associated with $ f(\cdot) $.
Then we could consider optimizing $\Theta$ and $\Phi$ simultaneously:
\begin{equation}
    \Theta^{*} =  \underset{\Theta, \Phi}{\mathrm{argmin}} \:
    \sum_{i} \sum_{j \in \mathcal{S}_{i}} \sum_{k \not \in \mathcal{S}_{i}} \mathcal{L}_{Ada} (i, j, k ; \Theta, \Phi) \,.
    \label{eq:optimize_simult}
\end{equation}
Unfortunately, directly minimizing the objective function as in
Eq.~\ref{eq:optimize_simult} does not achieve the desired purpose of generating
suitable adaptive margins. Since the margin-related term explicitly appears in
the loss function, constantly decreasing the value of the generated
margin is the straightforward way to reduce the loss. As a result all
generated margins have very small values or are set to zero, leading to
unsatisfactory results. In other words, the direct optimization of
$\mathcal{L}_{Ada} $ with respect to $ \Phi $ harms the optimization
of $ \Theta $.

\subsubsection{\bf Bilevel Optimization}
We model the learning of recommendation models and the generation of adaptive margins as a bilevel optimization problem~\cite{DBLP:journals/anor/ColsonMS07}:
\begin{equation}
\begin{aligned}
    & \underset{\Phi}{\mathrm{min}} \: \mathcal{J}_{outer}\left(\Theta^*(\Phi)\right):= \sum_{i} \sum_{j \in \mathcal{S}_{i}} \sum_{k \not \in \mathcal{S}_{i}} \mathcal{L}_{Fix} \left(i, j, k ; \Theta^{*}(\Phi)\right) \\
    & \mathrm{s.t.} \: \Theta^{*}(\Phi) = \underset{\Theta}{\mathrm{argmin}}\: \mathcal{J}_{inner}(\Theta, \Phi) := \sum_{i} \sum_{j \in \mathcal{S}_{i}} \sum_{k \not \in \mathcal{S}_{i}} \mathcal{L}_{Ada} (i, j, k ; \Theta, \Phi)\,.
\end{aligned}
\label{eq:bilevel}
\end{equation}
Here $ \Theta $ contains the model parameters $ \mathbf{\mu} $ and
$ \mathbf{\Sigma} $.  The objective function $\mathcal{J}_{inner}$
attempts to minimize $\mathcal{L}_{Ada}$ with respect to
$ \Theta $ while the objective function $\mathcal{J}_{outer}$
optimizes $\mathcal{L}_{Fix}$ with respect to $ \Phi $
through $ \Theta^*(\Phi) $. For simplicity, the $ m $ of
$\mathcal{L}_{Fix}$ in $\mathcal{J}_{outer}$ is set to $ 1 $ for
guiding the learning of $ f(\cdot; \Phi) $. Thus, we can have an alternating optimization to learn $ \Theta $ and $ \Phi $:
\begin{itemize}
    \item \textit{$ \Theta $ update phase} (Inner Optimization): Fix $ \Phi $ and optimize $ \Theta $.
    \item \textit{$ \Phi $ update phase} (Outer Optimization): Fix $ \Theta $ and optimize $ \Phi $.
\end{itemize}

\subsubsection{\bf Approximate Gradient Optimization}
As most existing models utilize gradient-based methods for optimization, a simple approximation strategy with less computation is introduced as follows:
\begin{equation}
    \begin{aligned}
    \nabla_{\Phi} \mathcal{J}_{outer}\left(\Theta^*(\Phi)\right)\approx
    \nabla_{\Phi} \mathcal{J}_{outer} \left(\Theta - \alpha \nabla_{\Theta} \mathcal{J}_{inner}(\Theta, \Phi) \right)\,.
    \end{aligned}
\end{equation}
In this expression, $\Theta$ denotes the current parameters including
$\mathbf{\mu}$ and $\mathbf{\Sigma}$, and $\alpha$ is the learning
rate for one step of inner optimization. Related approximations have been
validated in~\cite{DBLP:conf/wsdm/Rendle12,
  DBLP:conf/iclr/LiuSY19}. Thus, we can define a proxy function to link $\Phi$ with the outer optimization:
\begin{equation}
    \tilde{\Theta}(\Phi) := \Theta - \alpha \nabla_{\Theta} \mathcal{J}_{inner}(\Theta, \Phi)\, .
    \label{eq:gradient_approx}
\end{equation}

For simplicity, we use two optimizers $ \operatorname{OPT}_{\Theta} $ and $ \operatorname{OPT}_{\Phi} $ to update $ \Theta $ and $ \Phi $, respectively. The iterative procedure is shown in Alg.~\ref{alg:opt}.
\begin{algorithm}[hbt]
\SetAlgoLined
Initialize optimizers $\operatorname{OPT}_{\Theta}$ and $\operatorname{OPT}_{\Phi}$ \;
\While{not converged}{
$\Theta$ Update~(fix $\Phi^t$):\\ 
\Indp $\Theta^{t+1} \longleftarrow \operatorname{OPT}_{\Theta}\left(\Theta^t, \nabla_{\Theta^t} \mathcal{J}_{inner}(\Theta^t, \Phi^t)\right)$ \;
\Indm Proxy:\\
\Indp $\tilde\Theta^{t+1}(\Phi^t) := \Theta^t - \alpha \nabla_{\Theta^t} \mathcal{J}_{inner}(\Theta^t, \Phi^t)$ \;
\Indm$\Phi$ Update~(fix $\Theta^t$):\\
\Indp $\Phi^{t+1} \longleftarrow \operatorname{OPT}_{\Phi} \left( \Phi^t, \nabla_{\Phi^t}\mathcal{J}_{outer}\big(\tilde\Theta^{t+1}(\Phi^t)\big)\right)$ \;  
}
\caption{Iterative Optimization Procedure}
\label{alg:opt}
\end{algorithm}




\subsubsection{\bf The design of $ f(\cdot;\Phi) $}
We parameterize $ f(i,j,k;\Phi) $ with a neural network to generate the margin based on $ (i, j, k) $:
\begin{equation}
\begin{aligned}
    \mathbf{z}_{ijk} &= \mathrm{tanh}(\mathbf{W}_1 \cdot \mathbf{s}_{ijk} + \mathbf{b}_1) \,, \\
    m_{ijk} &= \mathrm{softplus}(\mathbf{W}_2 \cdot \mathbf{z}_{ijk} + \mathbf{b}_2) \,.
\end{aligned}
\end{equation}
Here $ \mathbf{W}_{*} $ and $ \mathbf{b}_{*} $ are learnable
parameters in $ f(\cdot;\Phi) $, $ \mathbf{s}_{ijk} $ is the input to
generate the margin, and $ m_{ijk} \in \mathbb{R} $ is the generated margin of $ (i, j, k) $. The activation function \textit{softplus}  guarantees $ m_{ijk} > 0 $. To promote a discrimative $ \mathbf{s}_{ijk} $ that reflects the relation between $ (i,j,k) $ and $ m_{ijk} $, the following form can be a fine-grained indicator:
\begin{equation}
\begin{aligned}
    \chi(\mathbf{u}_i, \mathbf{v}_j) &= [(u_{i,1} - v_{j,1})^2, (u_{i,2} - v_{j,2})^2,...,(u_{i,h} - v_{j,h})^2]^{\top} \\
    \mathbf{s}_{ijk} &= [\chi(\mathbf{u}_i, \mathbf{v}_j);\, \chi(\mathbf{u}_i, \mathbf{v}_k);\, \chi(\mathbf{u}_i, \mathbf{v}_k) \ominus \chi(\mathbf{u}_i, \mathbf{v}_j)]\,.
\end{aligned}
\label{eq:state_generation}
\end{equation}
Here $ \chi(\mathbf{u}_i, \mathbf{v}_j) \in \mathbb{R}^{h} $ is introduced to mimic the
calculation of Euclidean distance without summing over all
dimensions. $ \ominus $ denotes element-wise subtraction and
$ [...;...] $ denotes the concatenation operation. To improve the
robustness of $ f(\cdot; \Phi) $, we take as inputs the sampled
embeddings $ \mathbf{u}_{i} $ and $ \mathbf{v}_j $. To
perform backpropagation from $ \mathbf{u}_{i} $ and $ \mathbf{v}_j $,
we adopt the reparameterization
trick~\cite{DBLP:journals/corr/KingmaW13} for
Eq.~\ref{eq:gauss_distribution}:
\begin{equation}
\begin{aligned}
\mathbf{u}_{i} &= \mathbf{\mu}_{i}^{(U)} + (\mathbf{\Sigma}_{i}^{(U)})^{\frac{1}{2}} \odot \mathbf{\epsilon}_1 \,,\\
\mathbf{v}_{j} &= \mathbf{\mu}_{j}^{(I)} + (\mathbf{\Sigma}_{j}^{(I)})^{\frac{1}{2}} \odot \mathbf{\epsilon}_2 \,, 
\end{aligned}
\end{equation}
where $ \mathbf{\epsilon}_1, \mathbf{\epsilon}_2 \sim \mathcal{N}(\mathbf{0}, \mathbf{1}) $ and $ \odot $ is element-wise muliplication.


\subsection{User-User and Item-Item Relations} \label{sec:first_order_relation}
It is important to model the relationships between pairs of users or
pairs of items when developing recommender systems and strategies for
doing so effectively have been studied for many
years~\cite{DBLP:conf/www/SarwarKKR01,DBLP:conf/kdd/KabburNK13,DBLP:reference/sp/NingDK15}. For
example, item-based collaborative filtering methods use item rating
vectors to calculate the similarities between the items. Closely-related users or items may share the same interests or have similar attributes. For a certain user, items similar to the user's preferred items are potential recommendation candidates.

Despite this intuition, previous distance-based recommendation methods
\cite{DBLP:conf/www/HsiehYCLBE17,DBLP:conf/www/TayTH18} do not
explicitly take the user-user or item-item relationships into
consideration. As a result of relying primarily on user-item
information, the systems may fail to generate appropriate user-user or
item-item distances. To model the relationships between similar users or
items, we employ two ranking margin losses with adaptive margins to encourage similar users or items
to be mapped closer together in the latent space. 
Formally, the similarities between users or items are calculated from the user implicit feedback, which can be represented by a binary user-item interaction matrix.
We set a threshold on the calculated similarities to identify the similar users and items for a specific user $ i $ and item $ j $, respectively, denoted as $ \mathcal{N}^{U}_i $ and $ \mathcal{N}^{I}_j $.
We adopt the following losses for user pairs and item pairs, respectively:
\begin{align}
    &\left\{
    \begin{array}{rl}
       \mathcal{J}_{outer}^{U-U}
\: &:= \sum_{i} \sum_{p \in \mathcal{N}_i^{U}} \sum_{q \not\in \mathcal{N}_i^{U}} \mathcal{L}_{Fix} (i, p,q; \tilde\Theta^{t+1}_{U-U}) \,,\\
 \mathcal{J}_{inner}^{U-U}\: &:=   
 \sum_{i} \sum_{p \in \mathcal{N}_i^{U}} \sum_{q \not\in \mathcal{N}_i^{U}} \mathcal{L}_{Ada} (i,p,q ; \Theta^t, \Phi^t_{U-U}) \,,
    \end{array}
    \right.
\label{eq:hinge_user}\\
   &\left\{
    \begin{array}{rl}
       \mathcal{J}_{outer}^{I-I}
\: &:= \sum_{j} \sum_{p \in \mathcal{N}_j^{I}} \sum_{q \not\in \mathcal{N}_j^{I}} \mathcal{L}_{Fix} (j, p, q ; \tilde\Theta^{t+1}_{I-I}) \,,\\
 \mathcal{J}_{inner}^{I-I}\: &:=   
 \sum_{j} \sum_{p \in \mathcal{N}_j^{I}} \sum_{q \not\in \mathcal{N}_j^{I}} \mathcal{L}_{Ada} (j, p, q ; \Theta^t, \Phi^t_{I-I}) \,,
    \end{array} \right.
\label{eq:hinge_item}
\end{align}
where $ q $ is a randomly sampled user in Eq.~\ref{eq:hinge_user} and
a randomly sampled item in Eq.~\ref{eq:hinge_item}. $ U-U $ denotes
the user-user relation and $ I-I $ denotes the item-item relation. We
use $ \Phi^t_{U-U} $ and $ \Phi^t_{I-I} $ to update $
\Theta^{t+1}_{U-U} $ and $ \Theta^{t+1}_{I-I} $, respectively, which
are the same as in Alg.~\ref{alg:opt}.
We denote the indicator in Eq.~\ref{eq:state_generation} as $
\mathbf{s}_{ijk}^{U-I} $, then we generate $ \mathbf{s}_{ijq}^{U-U} $
and $ \mathbf{s}_{ijq}^{I-I} $ following the procedure described by
Eq.~\ref{eq:state_generation}.

\subsection{Model Training}
Let us denote the losses $ \mathcal{J}_{inner}^{U-I} $ and $ \mathcal{J}_{outer}^{U-I} $ to capture the interactions between users and items. Then we combine the loss functions presented in Section~\ref{sec:first_order_relation} to optimize the proposed model: 
\begin{equation}
\begin{aligned}
&\mathcal{J}_{inner} = \mathcal{J}_{inner}^{U-I} + \mathcal{J}_{inner}^{U-U} + \mathcal{J}_{inner}^{I-I} \,, \\ 
&\mathcal{J}_{outer} = \mathcal{J}_{outer}^{U-I} + \mathcal{J}_{outer}^{U-U} + \mathcal{J}_{outer}^{I-I} + \lambda ||\Phi||_{F}^{2}\,,
\end{aligned}
\label{eq:final_loss}
\end{equation}
where $ \lambda $ is a regularization parameter. We follow the same training scheme of Section~\ref{sec:adaptive_margin} to train Eq.~\ref{eq:final_loss}. To mitigate the curse of dimensionality issue~\cite{DBLP:conf/nips/BordesUGWY13} and prevent overfitting, we bound all the user/item embeddings within a unit sphere after each mini-batch training: $ ||\mathbf{\mu}|| \leqslant 1 $ and $ ||\mathbf{\Sigma}|| \leqslant 1 $. When minimizing the objective function, the partial derivatives with respect to all the parameters can be computed by gradient descent with back-propagation. 

\textbf{Recommendation Phase}. In the testing phase, for a certain user $ i $, we compute the distance $ \mathcal{W}_2(i, j)^{2} $ between user $ i $ and each item $ j $ in the dataset. Then the items that are not in the training set and have the shortest distances are recommended to user $ i $.

\section{Experiments} \label{sec:evaluation}

In this section, we evaluate the proposed model, comparing with the state-of-the-art methods on five real-world datasets.

\subsection{Datasets}

The proposed model is evaluated on five real-world datasets from various domains with different sparsities: \textit{Books}, \textit{Electronics} and \textit{CDs}~\cite{DBLP:conf/www/HeM16}, \textit{Comics}~\cite{DBLP:conf/recsys/WanM18} and \textit{Gowalla}~\cite{DBLP:conf/kdd/ChoML11}. The \textit{Books}, \textit{Electronics} and \textit{CDs} datasets are adopted from the Amazon review dataset with different categories, i.e., books, electronics and CDs. These datasets include a significant amount of user-item interaction data, e.g., user ratings and reviews. The \textit{Comics} dataset was collected in late 2017 from the \textit{GoodReads} website with different genres, and we use the genres of comics. The \textit{Gowalla} dataset was collected worldwide from the Gowalla website (a location-based social networking website) over the period from February 2009 to October 2010. In order to be consistent with the implicit feedback setting, we retain any ratings no less than four (out of five) as positive feedback and treat all other ratings as missing entries for all datasets. To filter noisy data, we only include users with at least ten ratings and items with at least five ratings. Table \ref{tab:data_statistics} shows the data statistics.

We employ five-fold cross-validation to evaluate the proposed model. For each user, the items she accessed are randomly split into five folds. We pick one fold each time as the ground truth for testing, and the remaining four folds constitute the training set. The average results over the five folds are reported.

\begin{table}[htbp]
\centering
\caption{\label{tab:data_statistics}The statistics of the datasets.}
\begin{tabular}{ |c|c|c|c|c|c| }
 \hline
 Dataset & \#Users & \#Items & \#Interactions & Density \\
 \hline
 \textit{Books} & 77,754 & 66,963 & 2,517,343 & 0.048\% \\ 
 \hline
 \textit{Electronics} & 40,358 & 28,147 & 524,906 & 0.046\% \\ 
 \hline
 \textit{CDs} & 24,934 & 24,634 & 478,048 & 0.079\% \\ 
 \hline
 \textit{Comics} & 37,633 & 39,623 & 2,504,498 & 0.168\% \\ 
 \hline
  \textit{Gowalla} & 64,404 & 72,871 & 1,237,869 & 0.034\% \\ 
 \hline
\end{tabular}
\vspace{-0.5cm}
\end{table}

\subsection{Evaluation Metrics}

We evaluate all models in terms of \textit{Recall@k} and \textit{NDCG@k}. For each user, Recall@k (R@k) indicates the percentage of her rated items that appear in the top $ k $ recommended items. NDCG@k (N@k) is the normalized discounted cumulative gain at $ k $, which takes the position of correctly recommended items into account. 

\subsection{Methods Studied}
To demonstrate the effectiveness of our model, we compare to the following recommendation methods.\\
\textit{Classical methods for implicit feedback}:
\begin{itemize}[leftmargin=*]
\item \textbf{BPRMF}, Bayesian Personalized Ranking-based Matrix Factorization~\cite{DBLP:conf/uai/RendleFGS09}, which is a classic method for learning pairwise personalized rankings from user implicit feedback.
\end{itemize}
\textit{Classical neural-based recommendation methods}:
\begin{itemize}[leftmargin=*]
\item \textbf{NCF}, Neural Collaborative Filtering~\cite{DBLP:conf/www/HeLZNHC17}, which combines the matrix factorization (MF) model with a multi-layer perceptron (MLP) to learn the user-item interaction function. 
\item \textbf{DeepAE}, the deep autoencoder \cite{DBLP:conf/cikm/MaZWL18}, which utilizes a three-hidden-layer autoencoder with a weighted loss function. 
\end{itemize}
\textit{State-of-the-art distance-based recommendation methods}:
\begin{itemize}[leftmargin=*]
\item \textbf{CML}, Collaborative Metric Learning~\cite{DBLP:conf/www/HsiehYCLBE17}, which learns a metric space to encode the user-item interactions and to implicitly capture the user-user and item-item similarities.
\item \textbf{LRML}, Latent Relational Metric Learning~\cite{DBLP:conf/www/TayTH18}, which exploits an attention-based memory-augmented neural architecture to model the relationships between users and items.
\item \textbf{TransCF}, Collaborative Translational Metric Learning~\cite{DBLP:conf/icdm/ParkKXY18}, which employs the neighborhood of users and items to construct translation vectors capturing the intensity of user–item relations.
\item \textbf{SML}, Symmetric Metric Learning with adaptive margin~\cite{DBLP:journals/corr/LiZZQZHH20}, which measures the trilateral relationship from both the user- and item-centric perspectives and learns adaptive margins.
\end{itemize}
\textit{The proposed method}:
\begin{itemize}[leftmargin=*]
\item \textbf{PMLAM}, the proposed model, which represents each user
  and item as Gaussian distributions to capture the uncertainties in user
  preferences and item properties, and incorporates an adaptive margin
  generation mechanism to generate the margins based on the sampled
  user-item triplets.
\end{itemize}

\subsection{Experiment Settings}
In the experiments, the latent dimension of all the models is set to
$50$ for a fair comparison. All the models adopt the same negative
sampling strategy with the proposed model, unless otherwise specified. For
BPRMF, 
the learning rate is set to $ 0.001 $ and the
regularization parameter is set to $ 0.001 $. With these parameters,
the model can achieve good results. For NCF, we follow the same model
structure as in the original paper~\cite{DBLP:conf/www/HeLZNHC17}. The
learning rate is set to $ 0.001 $ and the batch size is set to
$ 1024 $. For DeepAE, we adopt the same model structure employed in
the author-provided code and set the batch size to $ 512 $. The weight
of the positive items is selected from $ \{5, 10, 15, 20\} $ by a grid
search and the weights of all other items are set to $ 1 $ as
recommended in~\cite{DBLP:conf/icdm/HuKV08}. For CML, we use the
authors' implementation to set the margin to $ m=1 $ and the
regularization parameter to $ \lambda_{c}=1 $. For LRML, the learning
rate is set to $ 0.001 $, and the number of memories is selected from
$ \{ 5, 10, 20, 25, 50, 100 \} $ by a grid search. For TransCF, we
follow the settings in the original paper to select
$ \lambda,\, \lambda_{nbr},\, \lambda_{dist} \in \{0,\, 0.001,\,
0.01,\, 0.1\} $ and set the margin to $ 1 $ and batch size to
$ 1000 $, respectively. For SML, we follow the author's code to set
the user and item margin bound $ l $ to $ 1.0 $, $ \lambda $ to
$ 0.01 $ and $ \gamma $ to $ 10 $, respectively.

For our model, both the learning rate and $ \lambda $ are set to $ 0.001 $. For the $ Electronics $ and $ CDs $ datasets, we randomly sample $ 10 $ unobserved users or items as negative samples for each user and positive item.  This number is reduced to $ 2
$ for the other datasets to speed up the training process. The batch size is set to $ 5000 $ for all datasets. The dimension $ h $ is set to $ 50 $. The user and item embeddings are initialized by drawing each vector element independently from a zero-mean Gaussian distribution with a standard deviation of $ 0.01 $. Our experiments are conducted with PyTorch running on GPU machines (Nvidia Tesla P100).

\begin{table*}[ht]
\caption{\label{tab:performance_comparison}The performance comparison of all methods in terms of \textit{Recall@10} and \textit{NDCG@10}. The best performing method is boldfaced. The underlined number is the second best performing method. $ * $, $ ** $, $ *** $ indicate the statistical significance for $ p <= 0.05 $, $ p <= 0.01 $, and $ p <= 0.001 $, respectively, compared to the best baseline method based on the paired t-test. \textit{Improv.} denotes the improvement of our model over the best baseline method.}
\centering
\begin{tabular}{|c| c| c c| c c c c| l| c|}
\hline
& \textbf{BPRMF} & \textbf{NCF} & \textbf{DeepAE} & \textbf{CML} & \textbf{LRML} & \textbf{TransCF} & \textbf{SML} & \textbf{PMLAM} & \multicolumn{1}{l|}{\textbf{Improv.}} \\\hline
\multicolumn{10}{|c|}{Recall@10} \\
\hline

\textit{Books} & 0.0553  & 0.0568 & \underline{0.0817} & 0.0730 & 0.0565 & 0.0754 & 0.0581 & \textbf{0.0885**} & 8.32\% \\

\textit{Electronics} & 0.0243 & 0.0277 & 0.0253 & \underline{0.0395} & 0.0299 & 0.0353 & 0.0279 & \textbf{0.0469***} & 18.73\% \\

\textit{CDs} & 0.0730 & 0.0759 & 0.0736 & \underline{0.0922} & 0.0822 & 0.0851 & 0.0793 & \textbf{0.1129***} & 22.45\% \\

\textit{Comics} & 0.1966 & 0.2092 & \underline{0.2324} & 0.1934 & 0.1795 & 0.1967 & 0.1713 & \textbf{0.2417} & 4.00\% \\

\textit{Gowalla} & 0.0888 & 0.0895 & \underline{0.1113} & 0.0840 & 0.0935 & 0.0824 & 0.0894 & \textbf{0.1331***} & 19.58\% \\
\hline

\multicolumn{10}{|c|}{NDCG@10} \\ 
\hline
\textit{Books} & 0.0391  & 0.0404 & \underline{0.0590} & 0.0519 & 0.0383 & 0.0542 & 0.0415 & \textbf{0.0671**} & 13.72\% \\ 

\textit{Electronics} & 0.0111 & 0.0125 & 0.0134 & \underline{0.0178} & 0.0117 & 0.0148 & 0.0105 & \textbf{0.0234***} & 31.46\% \\

\textit{CDs} & 0.0383 & 0.0402 & 0.0411 & \underline{0.0502} & 0.0420 & 0.0461 & 0.0423 & \textbf{0.0619***} & 23.30\% \\

\textit{Comics} & 0.2247 & 0.2395 & \underline{0.2595} & 0.2239 & 0.1922 & 0.2341 & 0.1834 & \textbf{0.2753*} & 6.08\% \\

\textit{Gowalla} & 0.0806 & 0.0822 & \underline{0.0944} & 0.0611 & 0.0670 & 0.0611 & 0.0823 & \textbf{0.0984*} & 4.23\% \\
\hline
\end{tabular}
\vspace{-0.3cm}
\end{table*}

\subsection{Implementation Details}
To speed up the training process, we implement a two-phase sampling
strategy. We sample a number of candidates, e.g., 500, of negative
samples for each user every 20 epochs to form a candidate set. During
the next 20 epochs, the negative samples of each user are sampled from
her candidate set. This strategy can be implemented using multiple
processes to further reduce the training time.

Since none of the processed datasets has inherent user-user/item-item
information, we treat the user-item interaction as a user-item matrix
and compute the cosine similarity for the user and item pairs,
respectively~\cite{DBLP:conf/www/SarwarKKR01}. We set a threshold,
e.g., $ 0.2 $ on Amazon and Gowalla datasets and $ 0.4 $ on the Comics dataset, to
select the neighbors. These thresholds are chosen to ensure a reasonable degree of connectivity in the constructed graphs.

\begin{figure}[t!]
    \centering
    \begin{subfigure}[t]{0.25\textwidth}
        \centering
        \includegraphics[width=\linewidth]{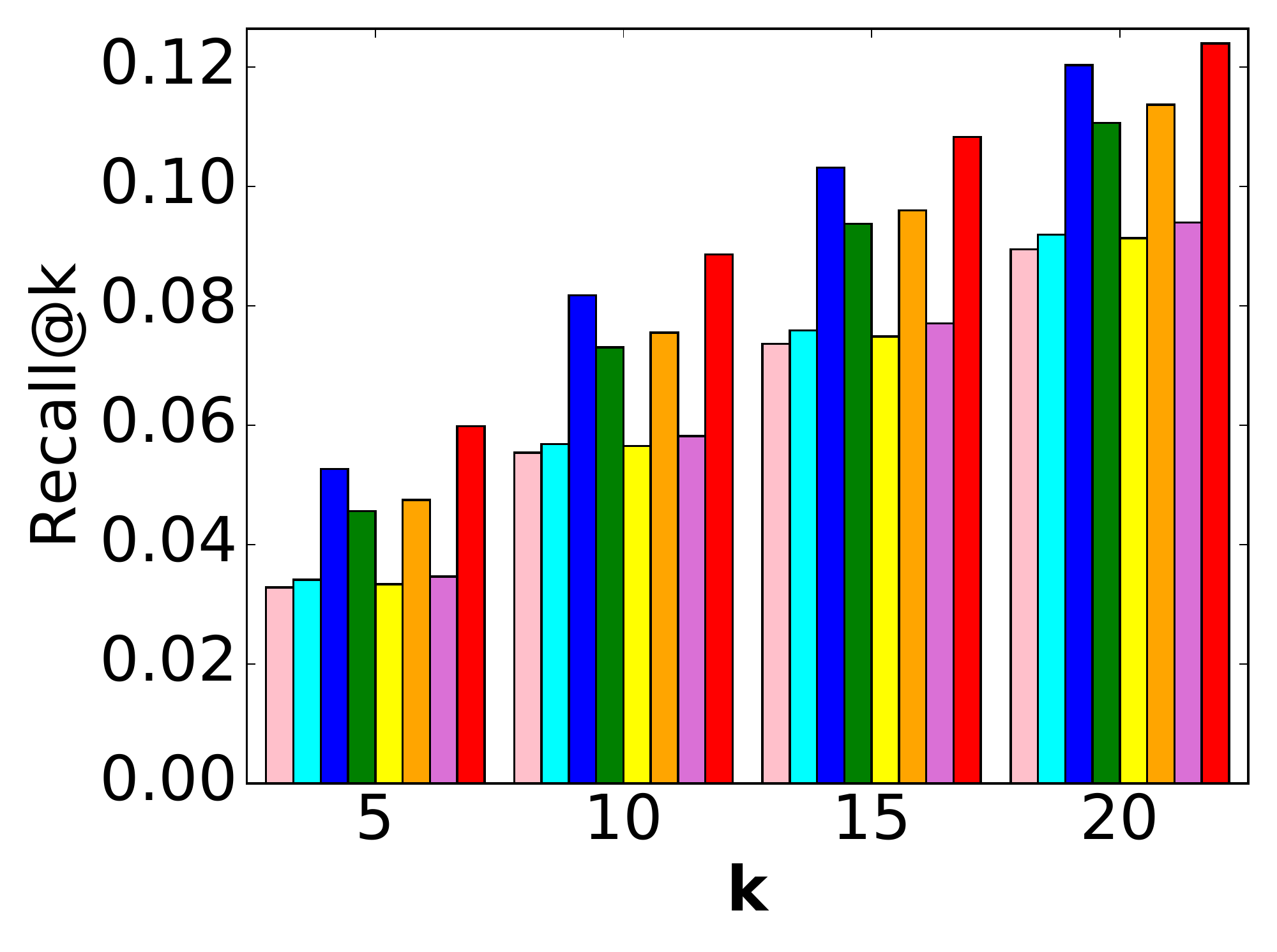}
        \caption{\label{fig:Books_recall} Recall@k on Books}
    \end{subfigure}%
    \begin{subfigure}[t]{0.25\textwidth}
        \centering
        \includegraphics[width=\linewidth]{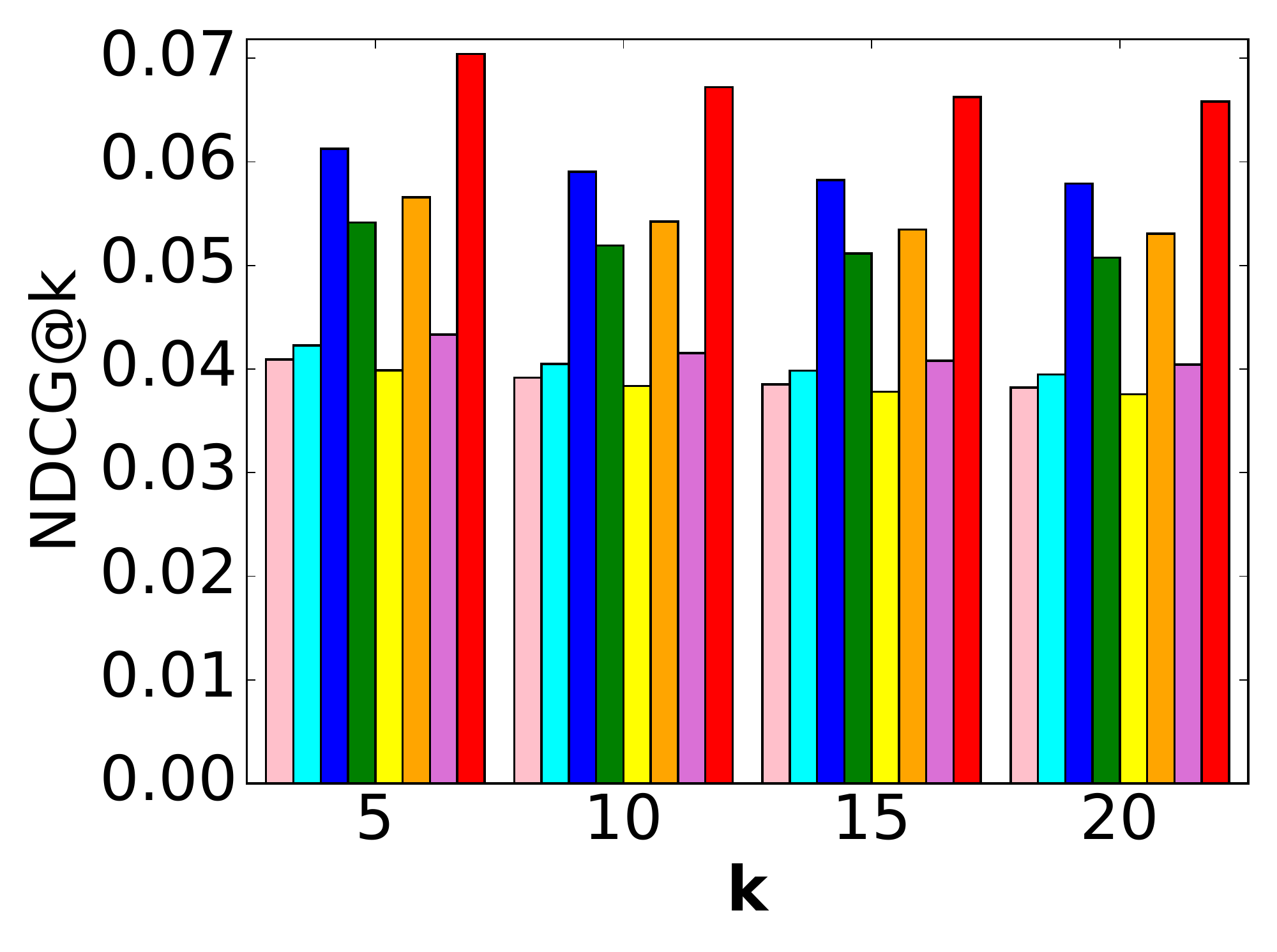}
        \caption{\label{fig:Books_ndcg} NDCG@k on Books}
    \end{subfigure}
    
    \begin{subfigure}[t]{0.25\textwidth}
        \centering
        \includegraphics[width=\linewidth]{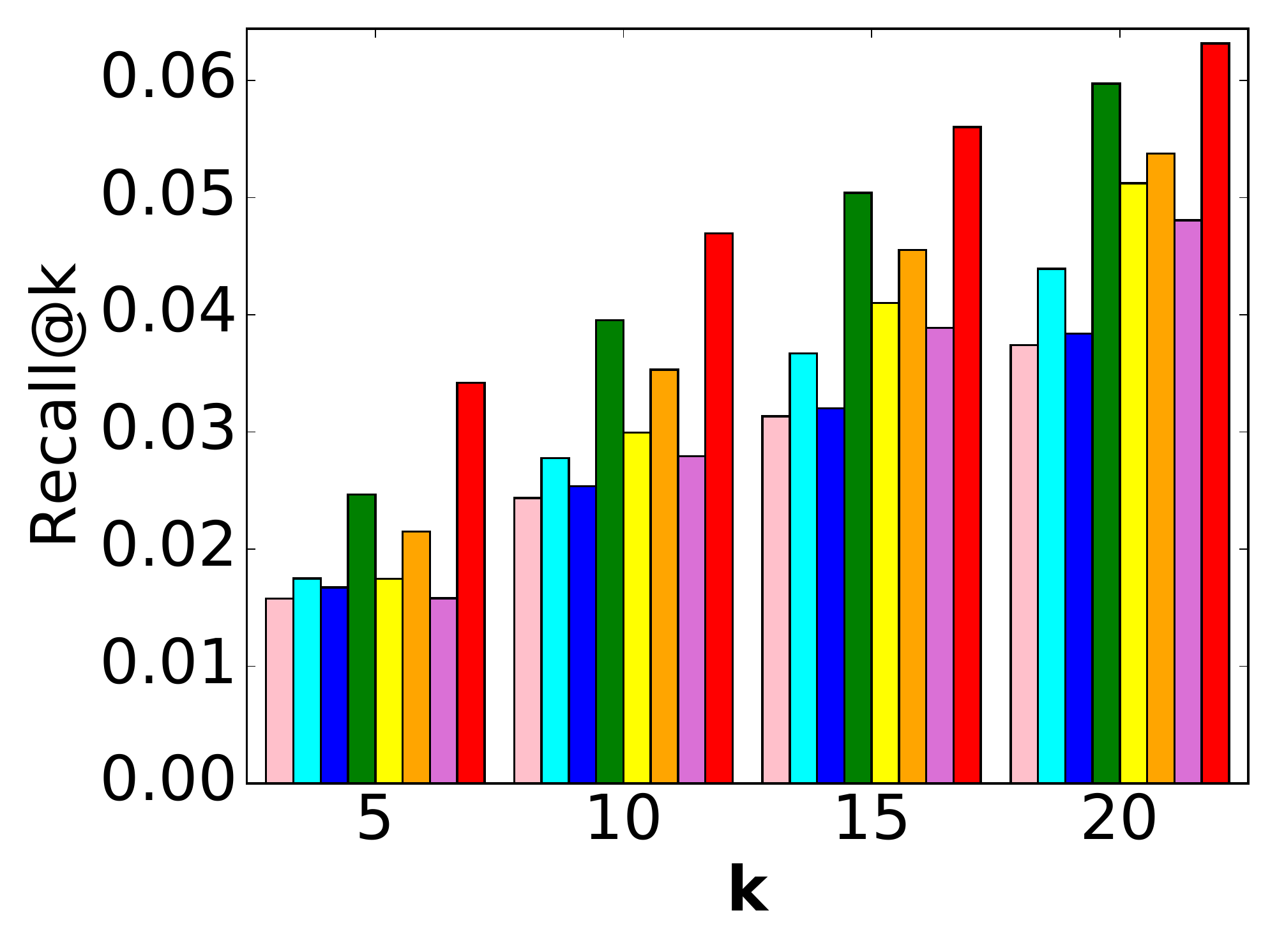}
        \caption{\label{fig:electronics_recall} Recall@k on Electronics}
    \end{subfigure}%
    \begin{subfigure}[t]{0.25\textwidth}
        \centering
        \includegraphics[width=\linewidth]{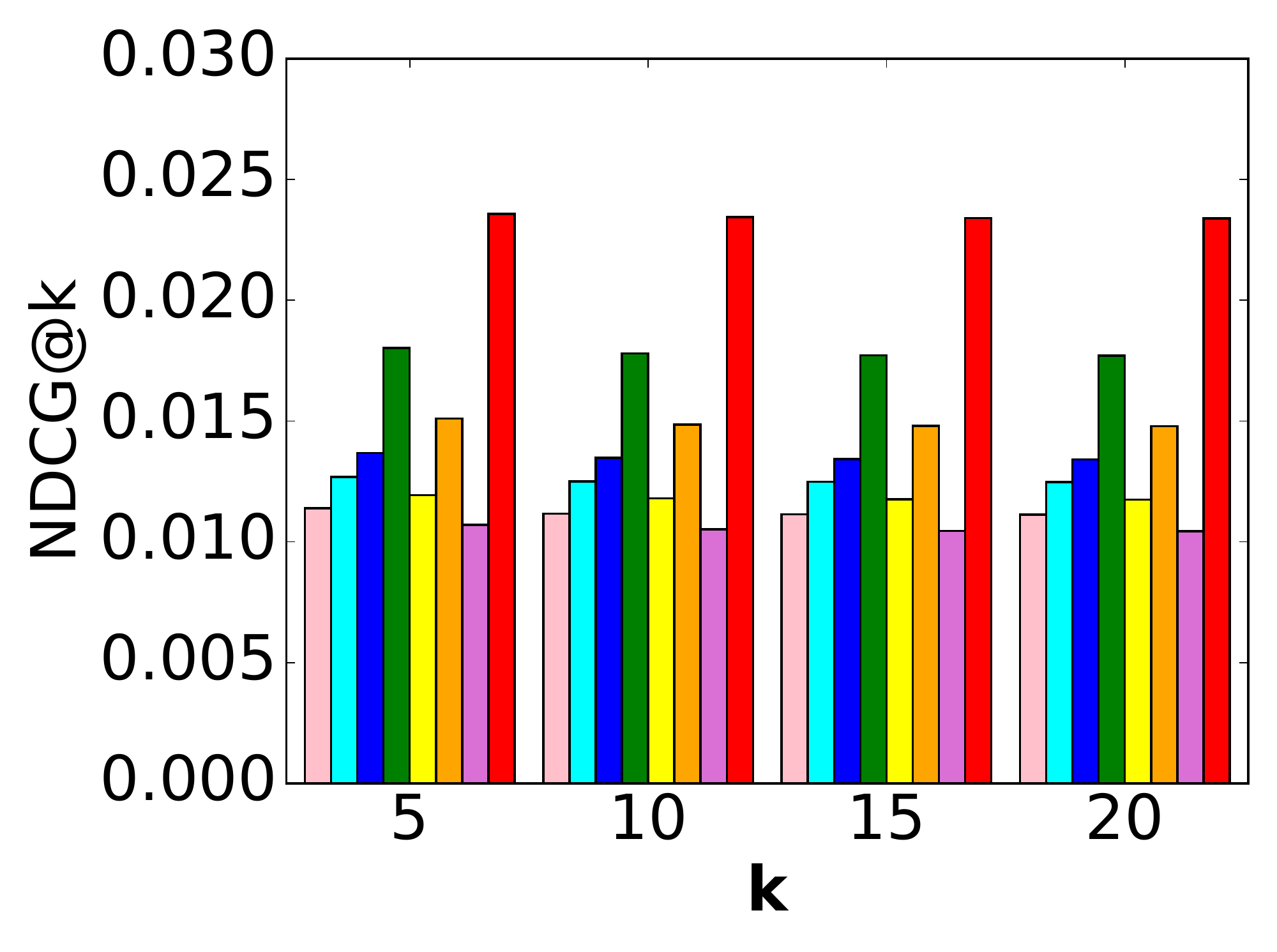}
        \caption{\label{fig:electronics_ndcg} NDCG@k on Electronics}
    \end{subfigure}
    
    \centering
    \begin{subfigure}[t]{0.25\textwidth}
        \centering
        \includegraphics[width=\linewidth]{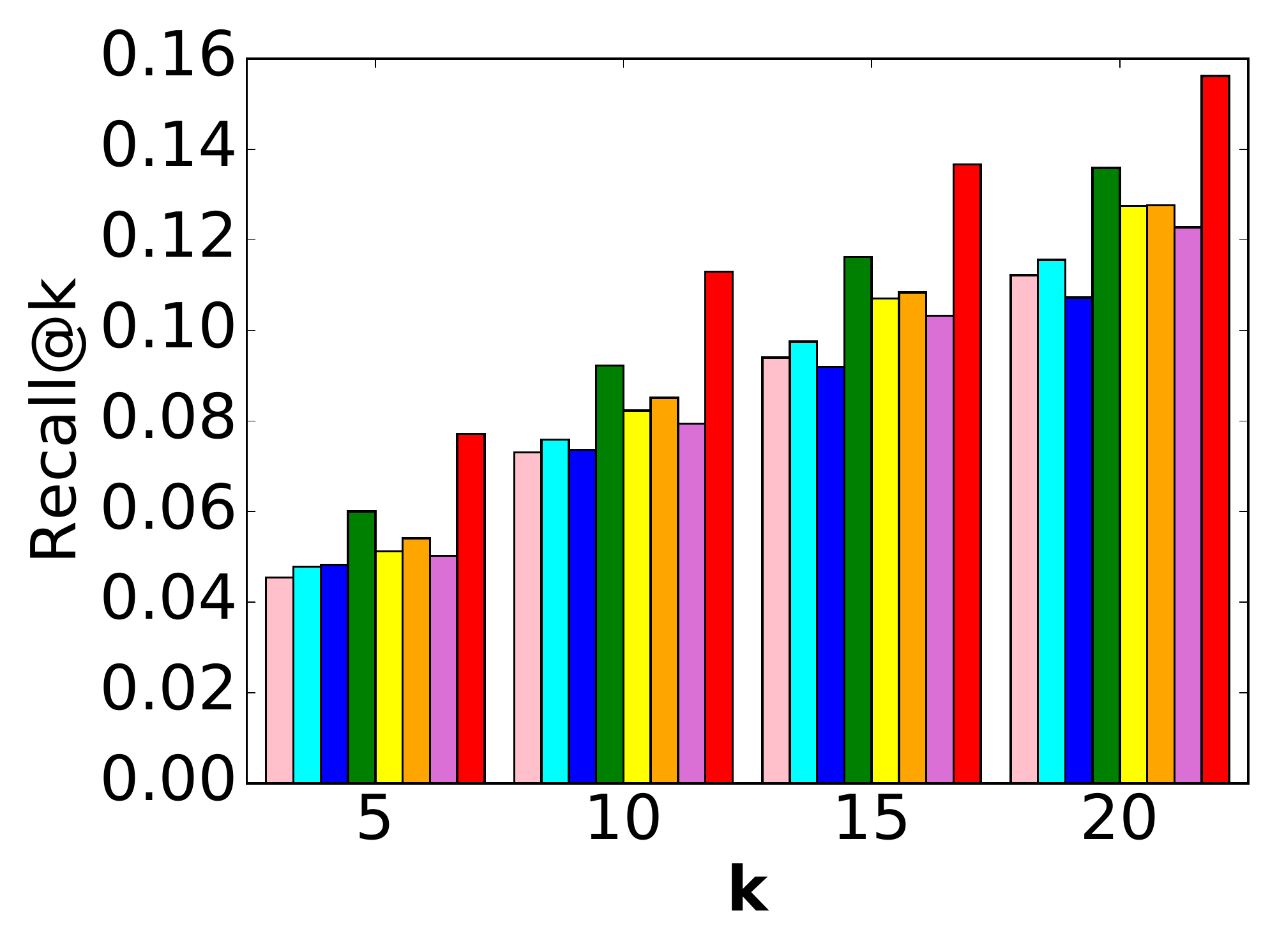}
        \caption{\label{fig:CDs_recall} Recall@k on CDs}
    \end{subfigure}%
    \begin{subfigure}[t]{0.25\textwidth}
        \centering
        \includegraphics[width=\linewidth]{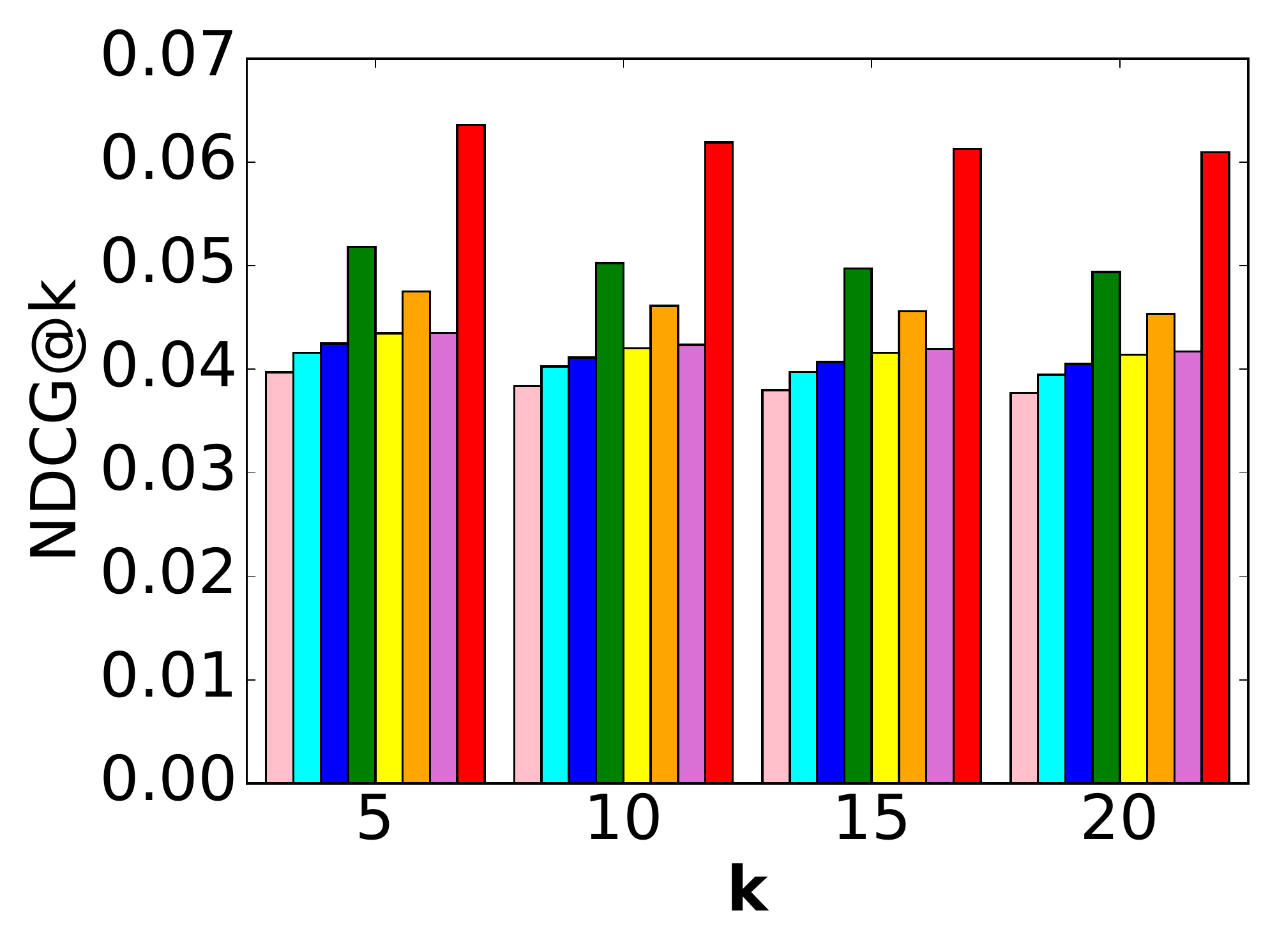}
        \caption{\label{fig:CDs_ndcg} NDCG@k on CDs}
    \end{subfigure}

    \begin{subfigure}[t]{0.25\textwidth}
    \centering
        \includegraphics[width=\linewidth]{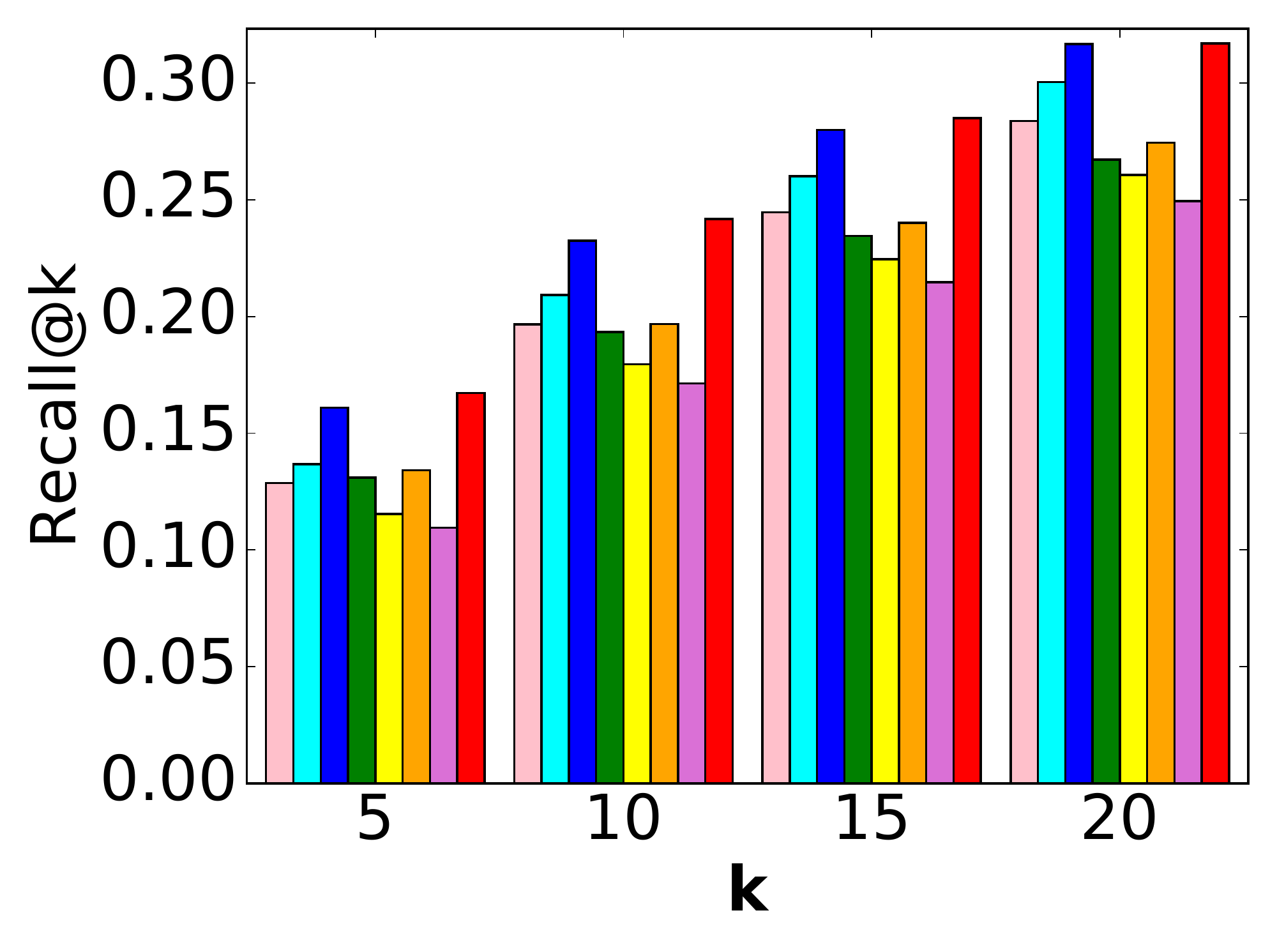}
        \caption{\label{fig:Comics_recall} Recall@k on Comics}
    \end{subfigure}%
    \begin{subfigure}[t]{0.25\textwidth}
    \centering
        \includegraphics[width=\linewidth]{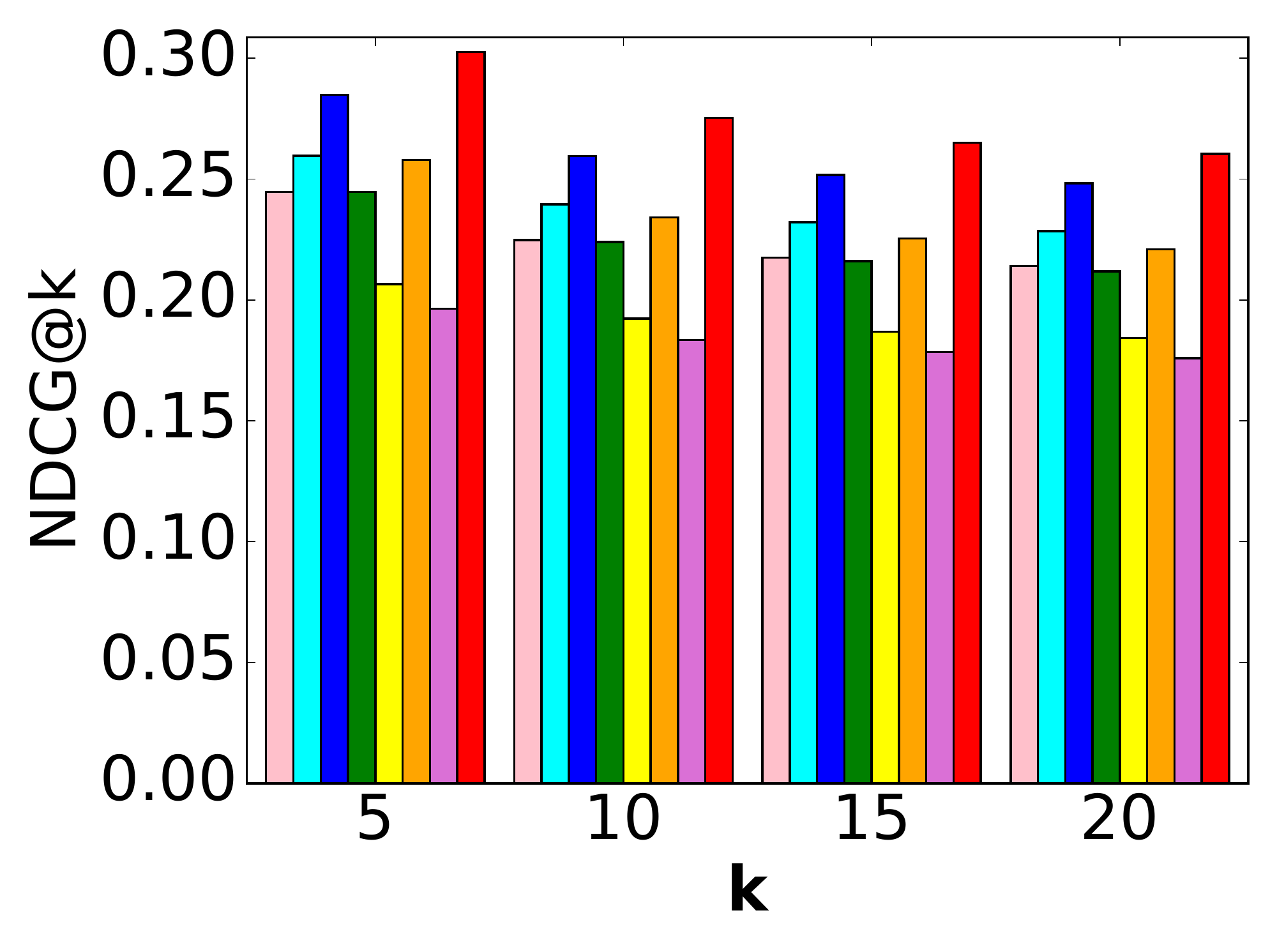}
        \caption{\label{fig:Comics_ndcg} NDCG@k on Comics}
    \end{subfigure}
    
    \begin{subfigure}[t]{0.25\textwidth}
    \centering
        \includegraphics[width=\linewidth]{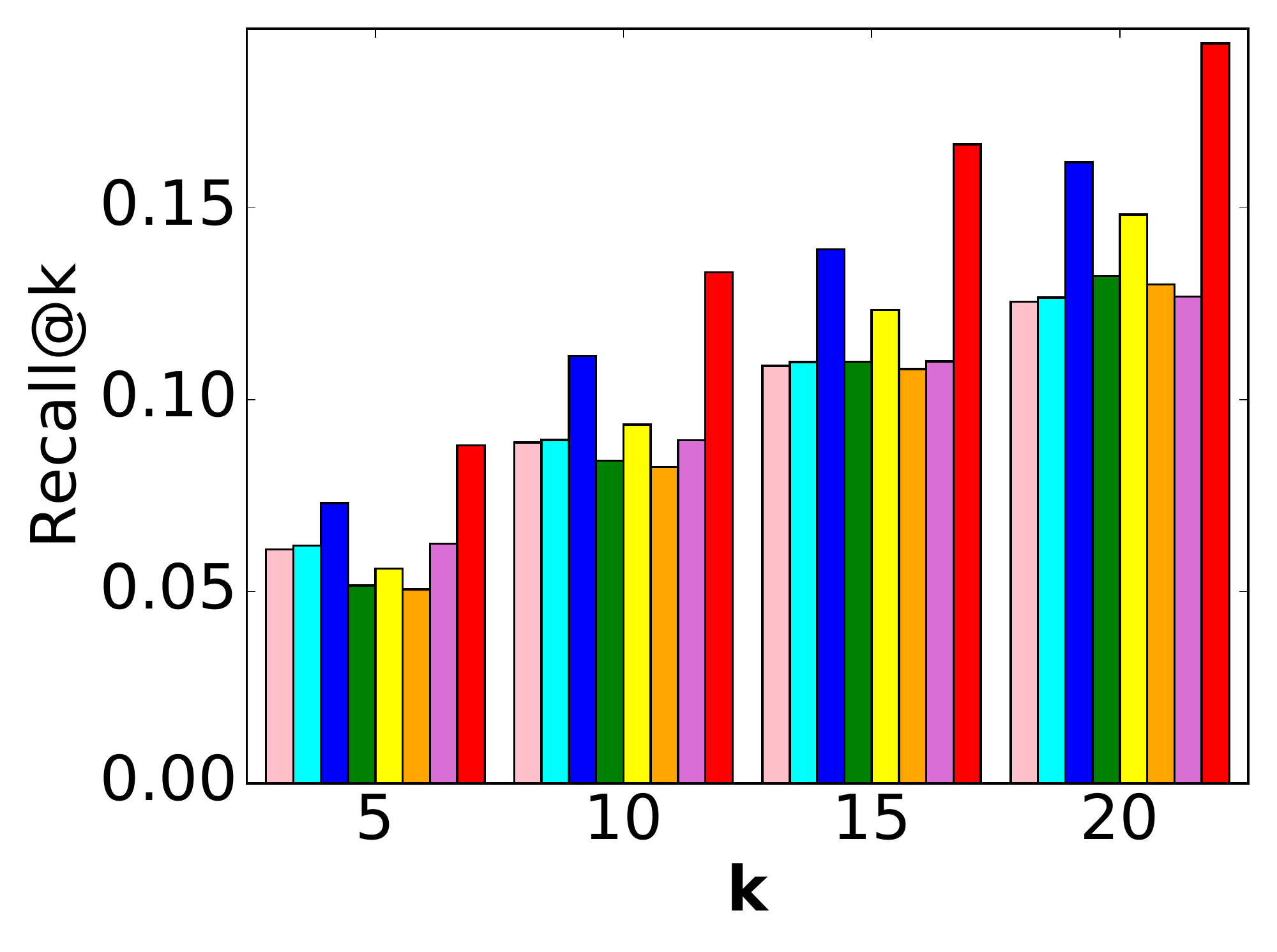}
        \caption{\label{fig:Children_recall} Recall@k on Gowalla}
    \end{subfigure}%
    \begin{subfigure}[t]{0.25\textwidth}
    \centering
        \includegraphics[width=\linewidth]{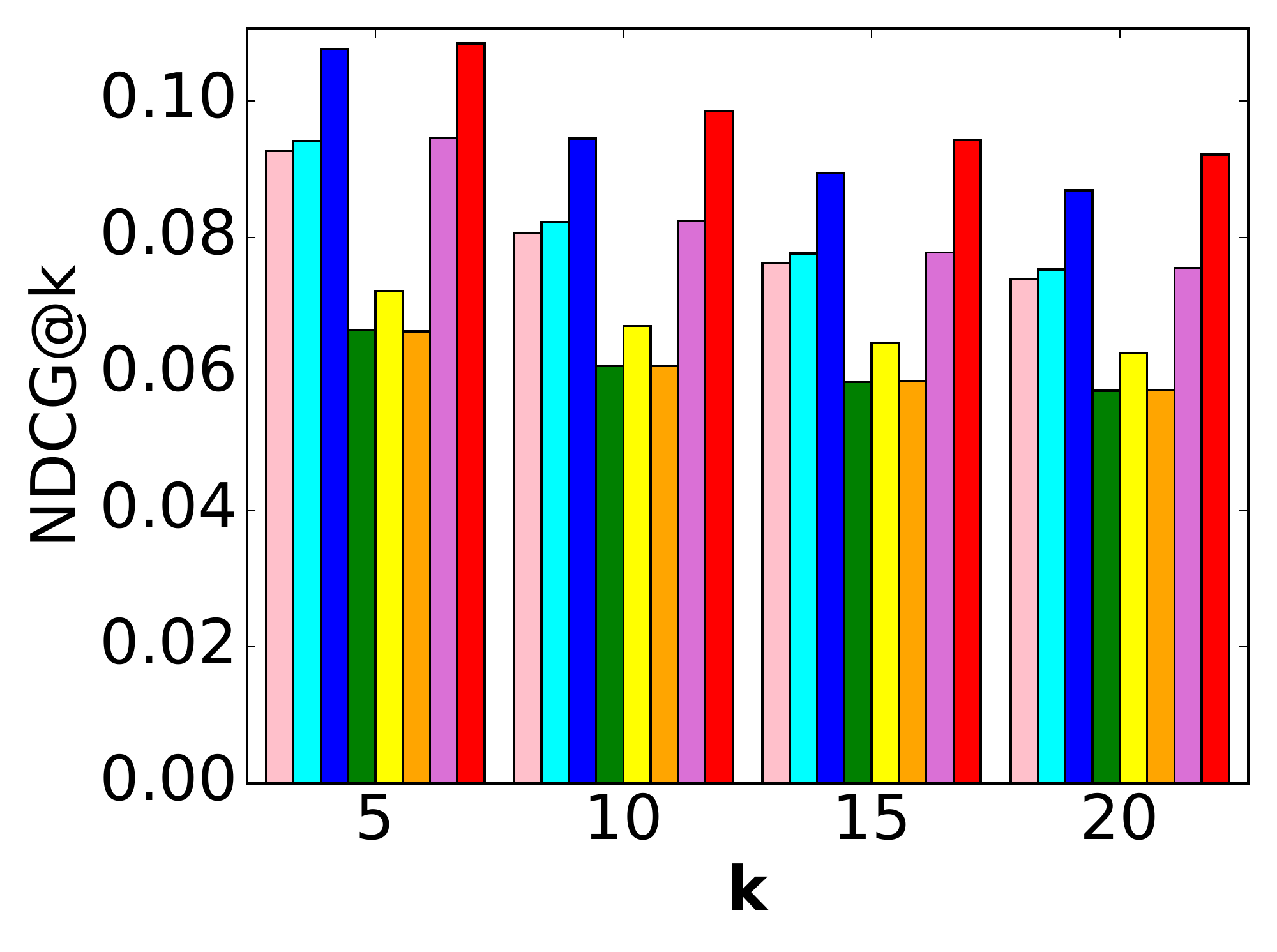}
        \caption{\label{fig:Children_ndcg} NDCG@k on Gowalla}
    \end{subfigure}
    
    \centering
    \begin{subfigure}[t]{0.45\textwidth}
    \centering
        \includegraphics[width=\linewidth]{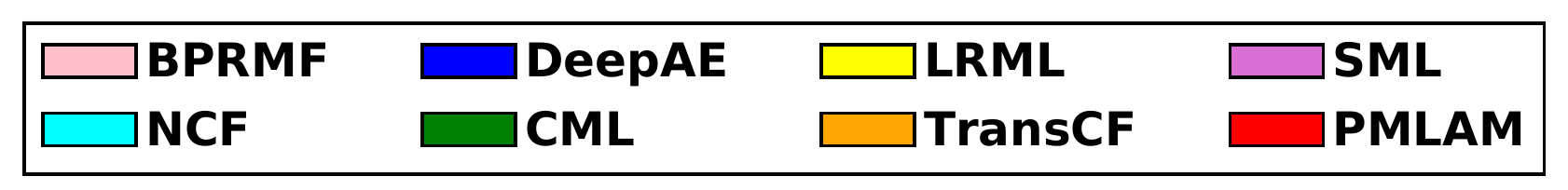}
    \end{subfigure}%
    \caption{\label{fig:performance_comparison}The performance comparison on all datasets.}
\vspace{-0.4cm}
\end{figure}

\subsection{Performance Comparison}
The performance comparison is shown in Figure~\ref{fig:performance_comparison} and Table~\ref{tab:performance_comparison}. Based on these results, we have several observations.

\textbf{Observations about our model}. {\em First}, the proposed model, PMLAM, achieves the best performance on all five datasets with both evaluation metrics, which illustrates the superiority of our model. \\
{\em Second}, PMLAM outperforms SML. Although SML has an adaptive
margin mechanism, it is achieved by having a learnable scalar margin
for each user and item and adding a regularization term to prevent the learned
margins from being too small. It can be challenging to
identify an appropriate regularization weight via hyperparameter tuning.
By contrast, PMLAM formulates the adaptive margin generation
as a bilevel optimization, avoiding the additional
regularization. PMLAM employs a neural network to
generate the adaptive margin, so the number of parameters related to margin generation does not increase with the number of users or items.\\
{\em Third}, PMLAM achieves better performance than TransCF. One major reason is that TransCF only considers the items rated by a user and the users who rated an item as the neighbors of the user and item, respectively, which neglects the user-user/item-item relations. PMLAM models the user-user/item-item relations by two margin ranking losses with adaptive margins.\\
{\em Fourth}, PMLAM makes better recommendations than CML and LRML.
These methods apply a fixed margin for all user-item triplets and
do not measure or model the uncertainty of learned user/item
embeddings. PMLAM represents each user and item as a Gaussian
distribution, where the uncertainties of learned user preferences and
item properties are captured by the covariance matrices.\\
{\em Fifth}, PMLAM outperforms NCF and DeepAE. These are MLP-based
recommendation methods with the ability to capture non-linear
user-item relationships, but they violate the triangle inequality when modeling user-item interaction. As a result, they can struggle to capture the fine-grained user preference for particular items~\cite{DBLP:conf/www/HsiehYCLBE17}. 

\textbf{Other observations}. {\em First}, all of the results reported
for the Comics dataset are
considerably better than those for the other datasets. The other four datasets are sparser and data sparsity negatively impacts recommendation performance.\\
{\em Second}, CML, LRML and TransCF perform better than SML on most of
the datasets. The adaptive margin regularization term in SML
struggles to adequately counterbalance SML's tendency to
reduce the loss by imposing small margins. Although it is reported
that SML outperforms CML, LRML and TransCF
in~\cite{DBLP:journals/corr/LiZZQZHH20}, the experiments are conducted
on three relatively small-scale datasets with only several thousands
of users and items. We experiment with much larger datasets;
identifying a successful regularization setting appears to be more
difficult as the number of users increases.\\
{\em Third}, TransCF outperforms LRML on most of the datasets. One
possible reason is that TransCF has a more effective translation
embedding learning mechanism, which incorporates the neighborhood information of users and items. TransCF also has a regularization term to further pull positive items closer to the anchor user. \\
{\em Fourth}, CML achieves better performance than LRML on most of the
datasets. CML integrates the weighted
approximate-rank pairwise (WARP) weighting
scheme~\cite{DBLP:journals/ml/WestonBU10} in the loss function to
penalize lower-ranked positive items. The comparison between CML and
LRML in~\cite{DBLP:conf/www/TayTH18} removes this component of CML.
The WARP scheme appears to play an important role in improving CML's performance.\\
{\em Fifth}, DeepAE outperforms NCF. The heuristic weighting function
of DeepAE can impose useful penalties to errors that occur during training when positive items are assigned lower prediction scores.

\begin{table}[ht]
\caption{\label{tab:ablation_analysis}The ablation analysis on the CDs and Electronics datasets. \textit{cat} denotes the concatenation operation and \textit{add} denotes the addition operation.}
\centering
\scalebox{0.95}{
\begin{tabular}{ |l|c|c|c|c| }
\hline
\multirow{2}{*}{Architecture} & \multicolumn{2}{c|}{\textit{CDs}} & \multicolumn{2}{c|}{\textit{Electronics}} \bigstrut \\ \cline{2-5} 
& R@10 & N@10 & R@10 & N@10 \bigstrut \\ 
\hline
(1) $ {Fix}^{U-I} $ + Deter\_Emb & 0.0721 & 0.0371 & 0.0241 & 0.0090 \\
(2) $ {Fix}^{U-I} $ + Gauss\_Emb & 0.0815 & 0.0434 & 0.0296 & 0.0110 \\ 
(3) $ {Ada}^{U-I} $ + Deter\_Emb & 0.0777 & 0.0415 & 0.0338 & 0.0125 \\ 
(4) $ {Ada}^{U-I} $-$cat$ + Deter\_Emb  & 0.0408 & 0.0204 & 0.0139 & 0.0055 \\ 
(5) $ {Ada}^{U-I} $-$add$ + Deter\_Emb & 0.0311 & 0.0158 & 0.0050 & 0.0018 \\ 
(6) $ {Ada}^{U-I} $ + Gauss\_Emb & 0.0856 & 0.0454 & 0.0365 & 0.0155 \\ 
(7) $ {Ada}^{U-I} $ + $ {Fix}^{U-U} $ + $ {Fix}^{I-I} $ & 0.0966 & 0.0526 & 0.0429 & 0.0189 \\ 
(8) PMLAM & \textbf{0.1129} & \textbf{0.0619} & \textbf{0.0469} & \textbf{0.0234} \\
\hline
\end{tabular}
}
\vspace{-0.4cm}
\end{table}

\subsection{Ablation Analysis} \label{sec:ablation}
To verify and
assess the relative effectiveness of the proposed user-item
interaction module, the adaptive margin generation module, and the
user-user/item-item relation module, we conduct an ablation study.
Table \ref{tab:ablation_analysis} reports the performance improvement
achieved by each module of the proposed model. Note that we compute Euclidean distances
between deterministic embeddings.
In (1), which serves as a baseline, we use the hinge loss with a fixed
margin (Eq.~\ref{eq:fixed_margin}) on deterministic embeddings of
users and items to capture the user-item interaction ($ m $ is set to
$ 1 $ which is commonly used
in~\cite{DBLP:conf/www/HsiehYCLBE17,DBLP:conf/www/TayTH18,DBLP:conf/icdm/ParkKXY18}).
In (2), as an alternative baseline, we apply the same hinge loss as in
(1), but replace the deterministic embeddings with parameterized
Gaussian distributions (Section~\ref{sec:wass}). In (3), we
use the adaptive margin generation module
(Section~\ref{sec:adaptive_margin}) to generate the margins for
deterministic embeddings. In (4), we concatenate the deterministic
embeddings of $ (i, j, k) $ to generate $ \mathbf{s}_{ijk} $ instead
of using Eq.~\ref{eq:state_generation}. In (5), we sum
the deterministic embeddings of $ (i, j, k) $ to generate
$ \mathbf{s}_{ijk} $ instead of using
Eq.~\ref{eq:state_generation}. In (6), we combine (2) and (3) to
generate the adaptive margins for Gaussian embeddings. In (7), we
augment (6) with user-user/item-item modeling but with a fixed margin, where the margin is also set to $ 1 $. In
(8), we add the user-user/item-item modeling with adaptive margins
(Section~\ref{sec:first_order_relation}) to replace the fixed margins
in the configuration of (7).

From the results in Table~\ref{tab:ablation_analysis}, we have several
observations. {\em First}, from (1) and (2), we observe that by
representing the user and item as Gaussian distributions and computing
the distance between Gaussian distributions, the performance improves.
This suggests that measuring the uncertainties of learned embeddings
is significant. {\em Second}, from (1) and (3) along with (2) and (6),
we observe that incorporating the adaptive margin generation module
improves performance, irrespective of whether deterministic or
Gaussian embeddings are used. These results demonstrate the
effectiveness of the proposed margin generation module. {\em Third},
from (3), (4) and (5), we observe that our designed inputs
(Eq.~\ref{eq:state_generation}) for margin generation facilitate the
production of appropriate margins compared to commonly used embedding
concatenation or summation operations. {\em Fourth}, from (2), (3) and
(6), we observe that (6) achieves better results than either (2) or
(3), demonstrating that Gaussian embedddings and adaptive margin
generation are compatible and can be combined to improve the model
performance. {\em Fifth}, compared to (6), we observe that the
inclusion of the user-user and item-item terms in the objective
function (7) leads to a large improvement in recommendation
performance. This demonstrates that explicit user-user/item-item
modeling is essential and can be an effective supplement to infer
user preferences. {\em Sixth}, from (7) and (8), we observe that adaptive margins
also improve the modelling of the user-user/item-item relations.

\begin{table}[ht]
\centering
\caption{\label{tab:adaptive_margin}A case study of the generated margin of sampled training triplets. The movie genre label is from the dataset.}
\scalebox{0.95}{
\begin{tabular}{ |c|l|l|c| }
 \hline
 User & Positive & Sampled Movie & Margin \\
 \hline
 \multirow{4}{*}{405} & \multirow{2}{*}{\textit{Scream} (Thriller)} & \textit{Four Rooms} (Thriller) & \textbf{1.2752} \\ \cline{3-4} 
                  & & \textit{Toy Story} (Animation) & 12.8004 \\ \cline{2-4} 
                  & \multirow{2}{*}{\textit{French Kiss} (Comedy)} & \textit{Addicted to Love} (Comedy) & \textbf{2.6448} \\ \cline{3-4} 
                  &  & \textit{Batman} (Action) & 12.4607 \\ \hline
 \multirow{4}{*}{66} & \multirow{2}{*}{\textit{Air Force One} (Action)} & \textit{GoldenEye} (Action) & \textbf{0.3216} \\ \cline{3-4} 
                  &  & \textit{Crumb} (Documentary) & 5.0010 \\ \cline{2-4} 
                  & \multirow{2}{*}{\textit{The Godfather} (Crime)} & \textit{The Godfather II} (Crime) & \textbf{0.0067} \\ \cline{3-4} 
                  &  & \textit{Terminator} (Sci-Fi) & 3.6335 \\ \hline
\end{tabular}
}
\vspace{-0.4cm}
\end{table}

\subsection{Case Study}
In this section, we conduct case studies to confirm whether the
adaptive margin generation can produce appropriate margins. To achieve
this purpose, we train our model on the MovieLens-100K dataset. This
dataset provides richer side information about movies (e.g., movie
genres), making it easier for us to illustrate the results. Since we
only focus on the adaptive margin generation, we use
deterministic embeddings of users and items to avoid the interference
of other modules. We randomly sample users from the dataset. For each
user, we sample one item that the user has accessed as the positive item
and two items the user did not access as negative items, where one
item has a similar genre with the positive item and the other does
not. The case study results are shown in
Table~\ref{tab:adaptive_margin}.

As shown in Table~\ref{tab:adaptive_margin}, our adaptive margin
generation module tends to generate a smaller margin value when the
negative movie has a similar genre with the positive movie, while
generating larger margins when they are distinct. The generated margins thus encourage the model to embed items with a higher probability of being preferred closer to the user's embedding.

\section{Conclusion}
In this paper, we propose a distance-based recommendation model for top-K recommendation. Each user and item in our model are represented by Gaussian distributions with learnable parameters to handle the uncertainties. By incorporating an adaptive margin scheme, our model can generate fine-grained margins for the training triples during the training procedure. To explicitly capture the user-user/item-item relations, we adopt two margin ranking losses with adaptive margins to force similar user and item pairs to map closer together in the latent space. Experimental results on five real-world datasets validate the performance of our model, demonstrating improved performance compared to many state-of-the-art methods and highlighting the effectiveness of the Gaussian embeddings and the adaptive margin generation scheme.  The code is available at \textcolor{blue}{\url{https://github.com/huawei-noah/noah-research/tree/master/PMLAM}}.

\bibliographystyle{ACM-Reference-Format.bst}
\bibliography{reference.bib}

\end{document}